\documentclass[screen,acmsmall]{acmart}

\AtBeginDocument{%
  \providecommand\BibTeX{{%
    \normalfont B\kern-0.5em{\scshape i\kern-0.25em b}\kern-0.8em\TeX}}}

\setcopyright{acmcopyright}

\acmJournal{TOSEM}
\acmVolume{}
\acmNumber{}
\acmArticle{}
\acmMonth{0}

\graphicspath{{./figs/}}
\DeclareGraphicsExtensions{.pdf,.jpeg,.png,.eps}

\usepackage{subcaption}
\usepackage{booktabs, tabularx}
\usepackage{xcolor}
\usepackage{algorithm}
\usepackage{algpseudocode}
\floatname{algorithm}{Procedure}

\usepackage{enumerate}
\usepackage{fancybox}  
\usepackage{multirow}
\usepackage{pifont}
\newcommand{\cmark}{\ding{51}}
\newcommand{\xmark}{\ding{55}}

\DeclareTextFontCommand{\emp}{\bfseries}

\usepackage[most]{tcolorbox}
\definecolor{custom-gray}{cmyk}{0, 0, 0, 0.7, 1.00}
\newtcolorbox{Summary}[2][]{
top=0.15in,
fonttitle=\bfseries,
colbacktitle=custom-gray,
colback=gray!5,
colframe=gray!40!black,
enhanced,
attach boxed title to top left={xshift=1.5em,yshift=-\tcboxedtitleheight/2},
boxed title style={size=small,colback=custom-gray},
drop shadow={black!50!white},
title=#2,#1}

\newcommand{\enquote}[1]{``#1''}
\newcommand{\etal}{et~al.~}

\newcommand{\verfmt}[1]{\textit{#1}}

\definecolor{mygreen}{RGB}{0,158,115}
\definecolor{myyellow}{RGB}{230,159,0}
\definecolor{myorange}{RGB}{255, 87, 37}
\definecolor{mycyan}{RGB}{0, 173, 181}
\definecolor{myblue}{RGB}{0, 114, 178}
\newcommand{\okcolor}[1]{\textbf{{#1}}}
\newcommand{\badcolor}[1]{{#1}}

\newcommand{\gopifix}[1]{}
\newcommand{\haofix}[1]{}

\begin{document}

\title{Studying the Impact of TensorFlow and PyTorch Bindings on Machine Learning Software Quality}

\author{Hao Li}
\email{li.hao@ualberta.ca}
\authornote{Hao Li and Cor-Paul Bezemer are with the Analytics of Software, GAmes And Repository Data (ASGAARD) Lab, University of Alberta, Canada.}
\orcid{0000-0003-4468-5972}
\affiliation{%
  \institution{University of Alberta}
  \city{Edmonton}
  \state{AB}
  \country{Canada}
  \postcode{T6G 2R3}
}

\author{Gopi Krishnan Rajbahadur}
\orcid{0000-0003-1812-5365}
\affiliation{%
	\institution{Centre for Software Excellence, Huawei Canada}
	\city{Kingston}
	\state{ON}
	\country{Canada}
	\postcode{K7L 1H3}}
\email{gopi.krishnan.rajbahadur1@huawei.com}

\author{Cor-Paul Bezemer}
\authornotemark[1]
\email{bezemer@ualberta.ca}
\orcid{0000-0002-0474-5718}
\affiliation{%
	\institution{University of Alberta}
	\city{Edmonton}
	\state{AB}
	\country{Canada}
	\postcode{T6G 2R3}
}

\begin{abstract}
Bindings for machine learning frameworks (such as TensorFlow and PyTorch) allow developers to integrate a framework's functionality using a programming language different from the framework's default language (usually Python). In this paper, we study the impact of using TensorFlow and PyTorch bindings in C\#, Rust, Python and JavaScript on the software quality in terms of correctness (training and test accuracy) and time cost (training and inference time) when training and performing inference on five widely used deep learning models. Our experiments show that a model can be trained in one binding and used for inference in another binding for the same framework without losing accuracy. Our study is the first to show that using a non-default binding can help improve machine learning software quality from the time cost perspective compared to the default Python binding while still achieving the same level of correctness. 

\end{abstract}

\ccsdesc[500]{Software and its engineering~Software libraries and repositories}
\ccsdesc[300]{Computing methodologies~Neural networks}
\ccsdesc[500]{Software and its engineering~Software performance}
\ccsdesc[500]{Software and its engineering~Correctness}

\keywords{Software engineering for machine learning, Software quality, Deep learning, Binding, TensorFlow, PyTorch
}


\maketitle

\newcommand{\rqone}{How do the studied bindings impact the training accuracy and test accuracy of the studied DL models?}
\newcommand{\rqtwo}{How do the studied bindings impact the cross-binding test accuracy of pre-trained models?}
\newcommand{\rqthree}{How do the studied bindings impact the training time of the studied DL models?}
\newcommand{\rqfour}{How do the studied bindings impact the inference time of pre-trained models?}

\newcommand{\motivation}{\emp{Motivation. }}
\newcommand{\approach}{\emp{Approach. }}
\newcommand{\findings}{\emp{Findings. }}
\newcommand{\runningexample}{\textbf{Running example. }}

\section{Introduction}\label{sec:introduction}



The rapidly improving capabilities of Deep Learning (DL) and Machine Learning (ML) frameworks have been the main drivers that allow new intelligent software applications, such as self-driving cars~\cite{gupta_sdc_2021, ni_sdc_2020} and robotic surgeons~\cite{shvets_robotic_2018, van_robotic_2021, esteva_robotic_2019}. These intelligent software systems all contain components that integrate one or more complex DL and/or ML algorithms. Fortunately, over the past decade, the need for coding these ML and DL algorithms from scratch has been largely eliminated by the availability of several mature ML frameworks and tools such as TensorFlow~\cite{abadi2016tensorflow} and PyTorch~\cite{pytorch}. These frameworks provide developers with a high-level interface to integrate ML functionality into their projects. Using such ML frameworks has several advantages including readily usable state-of-the-art algorithms, accelerated computing, and interactive visualization tools for data~\cite{nguyen_machine_2019}.

ML frameworks are typically accessed using Python, which is now the most popular programming language for ML applications~\cite{ben_braiek_open-closed_2018, nguyen_machine_2019, raschka_python_2020}. Gonzalez~\etal~\cite{gonzalez2020state} show that more than 56\% of the ML projects on GitHub are written in Python. However, many software projects do not use Python as their primary language\footnote{\url{https://githut.info}} and the developers of these projects might be unfamiliar with Python. Since learning a new language is a non-trivial task even for experienced developers~\cite{change_pl_2020}, these developers have to use a workaround to use the Python ML frameworks in their preferred programming language.

To help non-Python developers with the integration of an ML framework, 25\% of the popular ML frameworks offer one or more \emph{bindings} for other programming languages~\cite{mlbindings_2022}. These bindings expose the functionality of the framework in the binding's language. For example, TensorFlow provides a JavaScript binding\footnote{\url{https://github.com/tensorflow/tfjs}} that allows developers to integrate ML techniques directly in JavaScript. Because a binding adds an additional layer around the ML framework, it is important to investigate how the quality of the ML software created using these ML frameworks is impacted. For instance, different bindings may take different amounts of time to build a model.\footnote{As can be seen in this GitHub issue for TensorFlow: \url{https://github.com/tensorflow/tensorflow/issues/55476}} In addition, bugs in the bindings can introduce inconsistencies for trained models. For example, TensorFlow’s C\# binding had different results than the Python binding when loading an already trained model due to incorrectly handling `tf.keras.activations` functions.\footnote{\url{https://github.com/SciSharp/TensorFlow.NET/issues/991} and \url{https://github.com/SciSharp/TensorFlow.NET/pull/1001}} However, no one has systematically investigated the impact of using bindings for ML frameworks on the ML software quality; typically, studies focus on the software quality of the ML frameworks themselves~\cite{siebert_construction_2022, Chen_bugs_2022, liu_debts_2020}, or on the impact of the computing device on which the model executes~\cite{guo_empirical_2019}.

To illustrate the potential impact and importance of our study, consider the following real-world scenario. 
Anna's team uses JavaScript as the primary programming language. Since the team lacks ML or Python expertise, they collaborate with the company's ML team to integrate DL techniques into their projects. They are now considering using an ML framework's JavaScript binding for their project. However, they are concerned about how their developed ML software's quality is impacted by the binding; in particular, they are concerned about the correctness and time cost. 
There are three possible scenarios for integration of the binding that our study can assist with choosing the best option:

\begin{itemize}
    \item \textbf{Integration Scenario 1:} The ML team develops and trains the DL models and ships the pre-trained models to Anna. In this scenario, Anna needs to use the JavaScript binding to load the pre-trained models and perform model inference in her project. 
    \item \textbf{Integration Scenario 2:} The ML team assists Anna in training DL models in the project's native language which is JavaScript, allowing Anna to alter and maintain the code more efficiently. After training the DL models, Anna needs to deploy the trained models to the production environment in JavaScript as well. 
    \item \textbf{Integration Scenario 3:} Since computational resources for the project are very limited, Anna is also open to a third scenario, in which the ML team assists her in selecting the most efficient combination of training and inference bindings in any language. In this scenario, Anna is willing to hire an expert in the chosen language(s) to help with the integration of the binding(s) as long as the reduction in computational resources is large enough.
\end{itemize}

Therefore, in this paper we study the impact of bindings on two important ML software quality aspects:
\begin{itemize}
    \item \textbf{Correctness:} We evaluate if models trained using different bindings for a given ML framework have the same accuracy. We study (1)~training accuracy, which captures the model's classification performance on the train set during the training process, and (2)~test accuracy, which captures the classification performance of the final trained model on the test set. In addition, we measure whether the test accuracy is the same after loading a pre-trained model in a binding that was not used to train the model (the \emph{cross-binding} test accuracy). 
    \item \textbf{Time cost:} We evaluate if models trained using different bindings for an ML framework take similar time for training and making inferences. Bindings that produce models with a high time cost are expensive (in terms of computational resources), which limits their applicability. 
\end{itemize}

We conducted model training and model inference experiments using bindings for TensorFlow and PyTorch in C\#, Rust, Python, and JavaScript. In the model training experiments, we trained LeNet-1, LeNet-5, VGG-16, LSTM, GRU, and BERT models on the GPU in every binding (excluding BERT which is only trained on the Python bindings) using the same data and as far as possible, the same framework configuration. In the model inference experiments, we loaded pre-trained models and performed inference using every binding on the CPU and GPU. We do so to address the following research questions~(RQs), with RQ1 and RQ2 focusing on correctness, and RQ3 and RQ4 focusing on time cost:

\begin{enumerate}[\bfseries RQ1.]
\item \emp{\rqone}

During the training process, bindings for the same ML framework can have different training accuracies for the same model as well as varying test accuracy values~(2\% difference) in the final trained models.

\item \emp{\rqtwo}

The cross-binding test accuracy of the pre-trained models was not impacted by the bindings.

\item \emp{\rqthree}

Non-default bindings can be faster than the default Python bindings for ML frameworks. For instance, PyTorch's Python binding has the slowest training time for the studied models; PyTorch's C\# binding is more than two times faster than the Python binding in training the LeNet-5 model.

\item \emp{\rqfour}

Bindings can have very different inference times for the same pre-trained model, and the inference time of certain bindings on CPU can be faster than that of other bindings on GPU. For example, TensorFlow's Rust binding can perform inference faster for an LSTM model on CPU than the JavaScript binding on GPU~(73.9 vs. 177.7 seconds). 
\end{enumerate}

The main contributions of our paper are as follows:

\begin{enumerate}
    \item We are the first to study the impact of using different bindings for ML frameworks on the ML software quality in terms of correctness and time cost.
    \item We found that using a non-default binding can help improve ML software quality (from the time cost perspective) compared to the default Python binding of the studied frameworks in certain tasks, while still achieving the same level of correctness.
    \item We provide a replication package~\cite{replication_package}, which consists of the implementation of the studied ML models in the studied bindings, scripts for running the experiments, and Jupyter Notebooks for analyzing the experiment results. 
\end{enumerate}

The remainder of this paper is outlined as follows. Sections~\ref{sec:background} provides background information. Section~\ref{sec:exp_design} describes the design of our study. Sections~\ref{sec:correctness_eval} and~\ref{sec:cost_eval} present the results. Section~\ref{sec:implications} discusses the implications of our findings. Section~\ref{sec:relatedwork} gives an overview of related work. Section~\ref{sec:threadstovalidity} outlines threats to the validity of our study and Section~\ref{sec:conclusion} concludes the paper.

\section{Background}\label{sec:background}

\begin{figure}[t]
	\centering
	\includegraphics[width=0.7\textwidth]{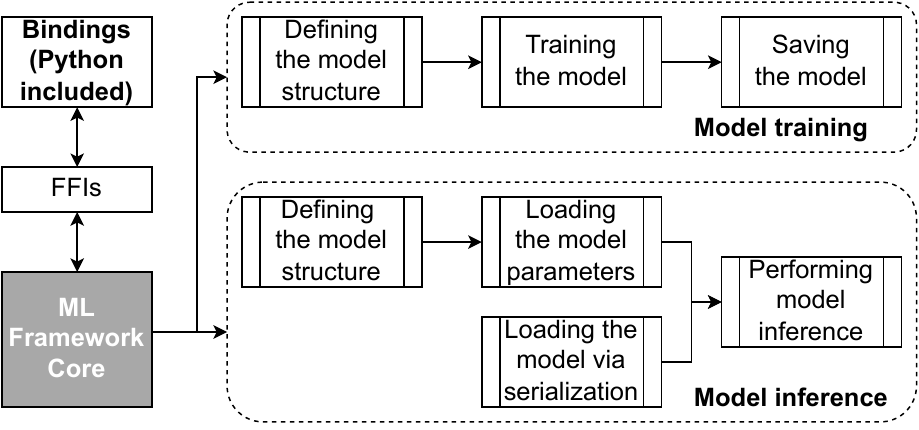}
	\caption{Bindings use the functionality of ML frameworks via foreign function interfaces~(FFIs) to train models and perform model inference.}
	\label{fig:background}
\end{figure}

\subsection{ML Frameworks}
Machine learning frameworks are software libraries that provide ML techniques to developers for the development and deployment of ML systems. Most popular ML frameworks are supported by large companies such as Google and Facebook~\cite{ben_braiek_open-closed_2018}. As shown in Figure~\ref{fig:background}, an ML framework provides interfaces to define the structure of a model, train the defined model using a selected optimizer, and save the trained model for later use. In addition, developers can deploy the trained models to the production environment by loading a saved (or \textit{pre-trained}) model and performing inference. ML frameworks can load a pre-trained model using (1)~the \textit{model parameters} (e.g., weights and hyperparameters) or (2)~\textit{serialization}. If only the model parameters are saved, developers first have to define the model structure before they can load the stored parameters into the defined model. When loading a serialized model, the ML framework can recreate the model from the saved file automatically since it contains both the structure and the weights of the pre-trained model.

Modern ML frameworks, such as TensorFlow and PyTorch, have been built upon a foundation that leverages parallel processing devices like GPUs. GPUs have proven to be highly efficient for tasks that demand parallel computation, especially in the realm of ML. Their architecture is inherently designed to handle multiple tasks simultaneously, allowing for massive parallelism. However, one significant characteristic of GPU computations that needs emphasis is their asynchronous nature. When a task is dispatched to a GPU, it does not always execute immediately. Instead, it often gets scheduled in a queue.\footnote{\url{https://developer.nvidia.com/blog/gpu-pro-tip-cuda-7-streams-simplify-concurrency/}} Consequently, a CPU might continue with its tasks believing that a GPU job is complete when, in fact, it has not even started. This asynchronous behaviour allows GPUs to optimize task execution but also necessitates careful synchronization when precise timing or task ordering is crucial.

\subsection{Bindings for the ML frameworks}
Python is the most popular programming language for ML applications~\cite{ben_braiek_open-closed_2018, nguyen_machine_2019}, but developers in other languages also have the need for using ML algorithms. Developers might choose an existing ML framework in their preferred language or they have to create a new one from scratch~(which requires a large amount of work and is error-prone). Another alternative is to use a \textit{binding} in their preferred language, which provides interfaces to the functionality of an existing ML framework in the language of the binding~\cite{mlbindings_2022}. 

As shown in Figure~\ref{fig:background}, bindings access the functionality of the ML framework through foreign function interfaces~(FFIs) without recoding the library. FFIs bridge the gap between programming languages, allowing developers to reuse code from other languages. For example, TensorFlow's Rust binding\footnote{\url{https://github.com/tensorflow/rust/tree/master/tensorflow-sys}} uses the FFI provided by the Rust language\footnote{\url{https://doc.rust-lang.org/rust-by-example/std_misc/ffi.html}} to access TensorFlow functionality. Since the GPU support is provided by the underlying C/C++ computational core of ML frameworks, bindings typically leverage FFIs to access these functionalities. For example, the Python bindings for TensorFlow and PyTorch make use of SWIG\footnote{\url{https://www.swig.org/}} (Simplified Wrapper and Interface Generator) and Pybind11\footnote{\url{https://github.com/pybind/pybind11}} to generate FFIs for its Python binding to tap into the C++ backend which includes the ability to access the GPU. However, the efficiency in leveraging GPU resources may vary among different bindings.

\begin{figure*}[t]
	\centering
	\includegraphics[width=\textwidth]{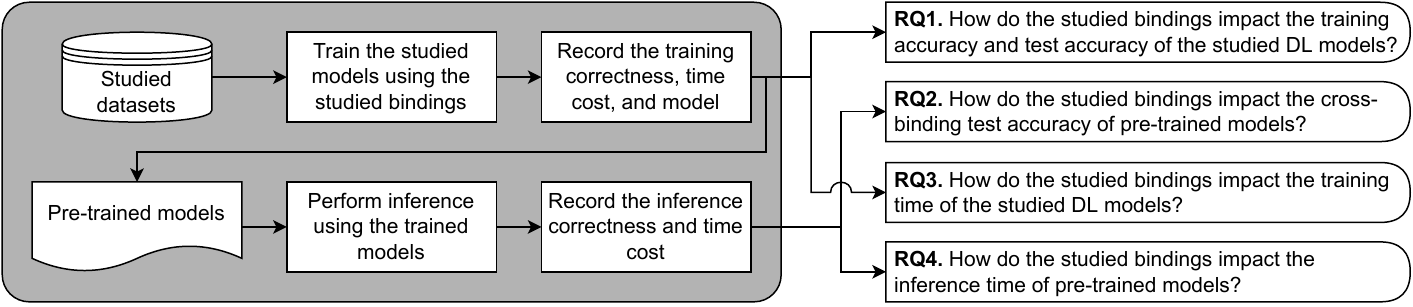}
	\caption{Overview of the study design.}
	\label{fig:methodology}
\end{figure*}

\section{Study Design}\label{sec:exp_design}

In this section, we first describe our experimental environment and the studied datasets, models, ML frameworks, and bindings. Then, we discuss how we evaluate the correctness and time cost in the model training and model inference experiments. Finally, we introduce the experimental setup of our study. Figure~\ref{fig:methodology} gives an overview of our study design.

\begin{table}[t]
	\centering	
	\caption{Our studied datasets and models. (Each model is paired with a dataset for the experiments)}
	\label{tab:datasets_models}
	\begin{tabular}{lrrllr}
		\toprule
		\multirow{2}{*}{\textbf{Dataset}} & \multicolumn{2}{c}{\textbf{\#Samples$^*$}}               &  & \multicolumn{2}{c}{\textbf{Model}} \\ \cmidrule{2-3} \cmidrule{5-6} 
		& \multicolumn{1}{c}{\textbf{Train}} & \multicolumn{1}{c}{\textbf{Test}} &  & \textbf{Name}      & \textbf{\#Parameters}  \\ \midrule
		\multirow{2}{*}{MNIST}    & \multirow{2}{*}{60,000} & \multirow{2}{*}{10,000} &  & LeNet-1 & 4,326      \\
		&      &      &  & LeNet-5   & 61,706        \\ \midrule
		CIFAR-10 & 50,000 & 10,000 &  & VGG-16  & 33,650,890 \\ \midrule
		\multirow{2}{*}{IMDb}     & \multirow{2}{*}{25,000} & \multirow{2}{*}{25,000} &  & LSTM    & 4,665,537  \\
		&        &        &  & GRU     & 4,250,817  \\  \midrule
		SQuAD & 87,599 & 10,570 &  & BERT (base)  & 108,893,186  \\ \bottomrule
\multicolumn{6}{l}{%
    \footnotesize{$^*$ The split of the training and test set is provided by the dataset.}
} 
\end{tabular}
\end{table}

\subsection{Environment setting}

We set up our experimental environment on a dedicated laboratory server provided by ISAIC\footnote{\url{https://isaic.ca/}}, where we can control the execution of other running tasks. The server runs Ubuntu Linux \verfmt{20.04} with Linux kernel \verfmt{5.11.0}. We used the CUDA \verfmt{11.1.74} and cuDNN \verfmt{8.1.0} GPU-related libraries. The hardware specifications of the server are as follows:

\begin{itemize}
	\item GPU: 2x NVIDIA TU102 [TITAN RTX] (24 GB)
	\item CPU: 3.30 GHz Intel(R) Core(TM) i9-9820X
	\item RAM: 100 GB
\end{itemize}

\subsection{Studied datasets and models}

Table~\ref{tab:datasets_models} presents the datasets and models used in this study, specifically pairing each model with the dataset used in the experiments. The datasets we studied are MNIST~\cite{mnist_dataset}, CIFAR-10~\cite{cifar_dataset}, IMDb review~\cite{imdb_dataset}, and SQuAD~\cite{rajpurkar_squad_2016}. These datasets are widely used as benchmarks in ML research~\cite{guo_empirical_2019, dataset_model_hu_2022, dataset_serhat_2021, dataset_model_li_2022, dataset_lin_2020, nguyen_machine_2019, cvdataset_wang_2022, dataset_xu_2021}. The models we studied are LeNet~\cite{lenet}, VGG~\cite{vgg16}, LSTM~\cite{lstm}, GRU~\cite{gru}, and BERT (the base model)~\cite{bert_2019} as all of them are typically paired with these datasets in various research domains~\cite{dataset_model_abdullahi_2021, dataset_model_frankle_2020, guo_empirical_2019, dataset_model_hourrane_2019, dataset_model_hu_2022, cvdataset_wang_2022, dataset_model_yan_2021, wang_bert_2020, ciborowska_bert_2022, yang_bert_2023, shen_bert_2023}.

MNIST and CIFAR-10 are datasets for image classification tasks. MNIST contains 70,000 grayscale images of handwritten digits, serving as a benchmark for evaluating classification models like LeNet-1 and LeNet-5. We used the CIFAR-10 dataset, which contains 60,000 colour images of 10 different objects, to train the VGG-16 model. The primary metric for these classification tasks is accuracy, reflecting the proportion of correctly identified images out of the total dataset.

The IMDb review dataset is utilized for sentiment analysis (text classification). The dataset contains 25,000 positive and 25,000 negative text reviews of movies. We used it to train the LSTM and GRU models to analyze the sequential nature of text data. Both LSTM and GRU models utilize a recurrent neural network (RNN) structure for handling sequential data, and we integrated a word embedding~\cite{word_embedding_collobert_2008} on the IMDb dataset in our experiments. The performance is measured by accuracy which indicates the model's ability to correctly classify reviews.

SQuAD is a dataset for the extractive question-answering task. SQuAD contains around 100,000 question-answer pairs, where the questions are posed by crowdworkers on a set of Wikipedia articles and the answer to every question is a text span from the corresponding reading passage. We used SQuAD to train the BERT-base model, leveraging the model's capability in language understanding. The task is to identify the exact text span (i.e., start and end positions) within the given passage that answers a question. The evaluation metric for SQuAD is the exact match score~\cite{rajpurkar_squad_2016}, which calculates the percentage of questions for which the model's answer exactly matches the annotated answer.


\subsection{Studied ML frameworks}

We study the latest stable versions (at the time of starting our study) of TensorFlow~\cite{abadi2016tensorflow}~(\verfmt{2.5.0}) and PyTorch\footnote{\url{https://github.com/pytorch/pytorch/releases/tag/v0.1.1}}~(\verfmt{1.9.0}), 
since they are two of the most popular ML frameworks. 
TensorFlow and PyTorch have recently grown in popularity as Caffe2 was merged into PyTorch in 2018\footnote{\url{https://caffe2.ai/}} and Keras became ``the high-level API of TensorFlow~2"~\cite{about_keras}.

\begin{table}[t]
\centering
\caption{Studied bindings for TensorFlow and PyTorch in software package ecosystems.}
\label{tab:bindings_info}
\begin{tabular}{llllrr}
\toprule
\multirow{2}{*}{\textbf{Framework}} & \multirow{2}{*}{\textbf{Name}} & \multirow{2}{*}{\textbf{Ecosystem}} & \multirow{2}{*}{\textbf{Language}} & \multirow{2}{*}{\textbf{Version}} & \multirow{2}{*}{\textbf{\# Stars$^\dagger$}} \\
                            &                       &       &              &        & \\ \midrule
\multirow{4}{*}{TensorFlow} & tensorflow            & PyPI  & Python       & 2.5.0  &  177,149 \\
                            & TensorFlow.NET        & NuGet & C\#          & 0.60.4 & 2,906 \\
                            & tensorflow            & Cargo & Rust         & 0.17.0 & 4,627 \\
                            & @tensorflow/tfjs-node & npm   & JavaScript$^*$ & 3.9.0  & 17,635 \\ \toprule
\multirow{4}{*}{PyTorch}    & pytorch               & PyPI  & Python       & 1.9.0  & 70,021 \\
                            & TorchSharp            & NuGet & C\#          & 0.93.9 & 946 \\
                            & tch                   & Cargo & Rust         & 0.5.0  & 3,178 \\
                            & @arition/torch-js     & npm   & JavaScript$^*$ & 0.12.3 & 252 \\ \bottomrule
\multicolumn{6}{l}{\footnotesize{$^*$ We wrote TypeScript code when using the JavaScript bindings.}} \\
\multicolumn{6}{l}{\footnotesize{$^\dagger$ The number of stars on GitHub recorded as of August 24, 2023.}}
\end{tabular}
\end{table}

\subsection{Studied Bindings}\label{sec:exp_design_studied_bindings}

The studied TensorFlow and PyTorch bindings are shown in Table~\ref{tab:bindings_info}. These bindings are all based on the same version of the studied ML frameworks~(i.e., TensorFlow~\verfmt{2.5.0} and PyTorch~\verfmt{1.9.0}). Notably, TensorFlow and PyTorch both utilize the Python bindings by default. The reason behind selecting bindings in these four software package ecosystems is twofold: (1)~Generally, PyPI (Python), npm (JavaScript), and NuGet (C\#) are the three most popular software package ecosystems for cross-ecosystem ML bindings~\cite{mlbindings_2022} and (2)~specifically, the Cargo ecosystem (Rust) is popular (according to the number of stars on GitHub) for both TensorFlow\footnote{\url{https://github.com/tensorflow/rust}} and PyTorch.\footnote{\url{https://github.com/LaurentMazare/tch-rs}} As shown in Table~\ref{tab:bindings_info}, the number of GitHub stars serves as a proxy for the popularity of a project in the software engineering domain~\cite{borges_stars_2016, han_stars_2019, fang_stars_2022, wolter_stars_2023, xia_stars_2023}, with TensorFlow's JavaScript binding being particularly notable. Although the number of stars for C\# and JavaScript bindings for PyTorch may appear low, we included these to ensure a fair comparison with TensorFlow bindings in respective ecosystems. 

\subsection{Correctness evaluation}

\textbf{Training correctness.} During the training process, the correctness is measured in each epoch using the training accuracy which is calculated by $Acc_{train} = {N_{correct}} / {N_{train}}$, 
where $N_{correct}$ is the number of correct predictions and $N_{train}$ is the number of data samples in the training set. 
For the final trained models, we use the test accuracy $Acc_{test} = {N_{correct}} / {N_{test}}$ as the evaluation metric for comparison, which is the accuracy on the test set.

\textbf{Inference correctness.} When we finish training a model, we use the test accuracy $Acc_{test}$ of this pre-trained model as a reference. Then, we perform inference with a studied binding for the pre-trained model on the test set to obtain the cross-binding test accuracy $Acc_{cross\_test} = {N_{correct}} / {N_{test}}$ using that binding. The difference between $Acc_{test}$ and $Acc_{cross\_test}$ is that the inference correctness is measured in the studied binding. For BERT on SQuAD, we use the exact match score~\cite{rajpurkar_squad_2016} instead of accuracy as the metric to evaluate the correctness.



\subsection{Time cost evaluation}

\textbf{Training time cost.} The training time cost measures the time spent training a model in seconds. Developers commonly train DL models on GPU rather than CPU since the training can be time-consuming and GPU can considerably shorten the training time~\cite{buber_gpu_2018, li_cpu_gpu_2019}. Hence, all model training experiments of bindings for ML frameworks are conducted on GPU and we measure the training time cost on GPU only.

\textbf{Inference time cost.} The inference time cost measures the time spent for performing inference with a pre-trained model on the test set in seconds. Since developers can deploy pre-trained models to a production environment which supports the CPU or GPU, the inference time cost of a binding is measured on both CPU and GPU.

\begin{table}[t]
\centering
\caption{Supported features of studied bindings for TensorFlow (TF) and PyTorch (PT).}
\label{tab:supported_features}
\begin{tabular}{lccccclcc}
\toprule
\multirow{2}{*}{} & \multirow{2}{*}{} & \multirow{2}{*}{\textbf{Training}} & \multicolumn{3}{c}{\textbf{Supported interfaces}} &  & \multicolumn{2}{c}{\textbf{Loading models}} \\ \cmidrule(lr){4-6} \cmidrule(l){8-9} 
                            &            &  & \textbf{CNNs} & \textbf{RNNs} & \textbf{BERT} &  & \textbf{Parameters} & \textbf{Serialization} \\ \midrule
\multirow{4}{*}{TF} & Python     &  \cmark    & \cmark & \cmark & \cmark &  & \cmark & \cmark  \\
                            & C\#        &  \cmark    & \cmark & \xmark$^\dagger$ & \xmark &  & \cmark & \xmark  \\
                            & Rust       &  \xmark$^*$ & \cmark & \cmark & \xmark &  & \xmark & \cmark  \\
                            & JavaScript &  \cmark    & \cmark & \cmark & \xmark &  & \xmark & \cmark   \\ \midrule
\multirow{4}{*}{PT}    & Python     &  \cmark    & \cmark & \cmark & \cmark &  & \cmark & \cmark  \\
                            & C\#        & \cmark    & \cmark & \cmark & \xmark &  & \cmark & \xmark  \\
                            & Rust       &  \cmark    & \cmark & \cmark & \xmark &  & \cmark & \cmark  \\
                            & JavaScript &  \xmark    & \xmark & \xmark & \xmark &  & \xmark & \cmark   \\ \bottomrule
\multicolumn{9}{l}{%
    \footnotesize{$^*$ Unlike other bindings, TensorFlow's Rust binding does not support the API (Keras-like) of TensorFlow 2.}
} \\
\multicolumn{9}{l}{\begin{tabular}[c]{@{}l@{}}
\footnotesize{$^\dagger$ TensorFlow's C\# binding has only recently introduced support for RNNs based on TensorFlow 2.10, however, our} \\ 
\footnotesize{study uses the C\# binding for TensorFlow 2.5.0 for consistency across all bindings.}
\end{tabular}}
\end{tabular}
\end{table}

\subsection{Experimental setup}\label{sec:exp_setup}

In this section, we detail our experimental setup with a running example of how we computed the correctness and time cost of LeNet-1 when trained and inferenced using the studied bindings for the studied ML frameworks.

\textbf{Step 1 -- Train the studied models using the studied bindings:}
We conduct model training experiments for each supported model-dataset pair~(as shown in Table~\ref{tab:datasets_models}). For a given model-dataset pair, each binding that supports the model's interface and training features~(as shown in Table~\ref{tab:supported_features}) trains the model from scratch on that dataset. For example, LeNet-1 and MNIST form one model-dataset pair and each supported binding trains LeNet-1 on MNIST independently. We repeat this process for each model-dataset pair in each binding that supports the model. For consistency, we ensure the following across all bindings for a given model-dataset pair:

\begin{itemize}
\item \textbf{Model structure.} 
We use interfaces that provide the same functionality in bindings to build up each layer of the studied models. However, not all bindings support model training, as indicated in Table~\ref{tab:supported_features}. As a result, we do not conduct training experiments with TensorFlow's Rust binding, PyTorch's JavaScript binding, and RNNs in TensorFlow's C\# binding.

\item \textbf{Training set and test set.} 
We use the provided split of the training set and test set from studied datasets. Before conducting experiments, we perform comprehensive data preprocessing, ensuring that all bindings can work with the same processed data across all experiments.

\item \textbf{Hyperparameters.}
We use the same hyperparameters~(e.g., the number of epochs and batch size) and optimizers from prior research~\cite{guo_empirical_2019}. However, TensorFlow's C\# binding does not support setting the momentum and weight decay hyperparameters for a stochastic gradient descent~(SGD) optimizer. Hence, we only set the learning rate for the SGD optimizer without enabling momentum and weight decaying when training the LeNet-1, LeNet-5, and VGG-16 models to maintain consistency across all bindings. In addition, to mitigate the risk of default hyperparameters influencing our results, we explicitly defined all configurable parameters and kept them the same across bindings.

\item \textbf{Random seed.} 
We fix the value of the random seed across bindings when training the same model to control the randomness.

\end{itemize}

In addition, we repeat the same training process five times for each binding with different random seeds (that are kept consistent across bindings) to reduce the impact of seed selection on the results. 

\runningexample We train the LeNet-1 model in TensorFlow's Python, C\#, and JavaScript bindings. These bindings all set the same random seed at the start of the training process. To build up the same convolution layers of the model, we use the ``Conv2D'' interface in Python, ``Conv2D'' in C\#, and the ``conv2d'' interface in JavaScript. In addition, we use SGD with a learning rate of 0.05 for all three bindings to train the LeNet-1 model.

\textbf{Step 2 -- Record the training correctness and save the model:}
We record the training accuracy in each epoch for all model training experiments. After the training is completed, we compute the trained model's test accuracy and save the model for later use. Considering the impact of randomness, we repeat the training process 5 times in each training experiment and analyze the distribution of the results to draw conclusions. 

\runningexample During training the LeNet-1 model in PyTorch's C\# binding, we calculate the training accuracy in each epoch and store the value. After finishing the training, we save the trained LeNet-1 model.

\begin{algorithm}[t]
\caption{Measuring Training Time Cost in PyTorch Bindings}\label{alg:training_time_tch}
\begin{algorithmic}[1]
\State $model, optimizer \gets \text{initModelAndOptimizer()}$ \Comment{Model and optimizer initialization}
\State $train\_set \gets \text{loadDataset()}$ \Comment{Load pre-processed training set}
\State $start \gets \text{getCurrentTime()}$ \Comment{Start the timer}
\For{$epoch \gets 1$ to $epochs$}
    \While{not \text{isEndOfDataset($trainSet$)}}
        \State $inputs, labels \gets \text{getNextBatch($train\_set$)}$ \Comment{Batch data loading$^{*1}$}
        \State $outputs \gets model(inputs)$ \Comment{Start forward propagation$^{*2a}$}
        \State $loss \gets \text{calculateLoss($outputs, labels$)}$ \Comment{Loss calculation$^{*2b}$}
        \State $loss.backward()$ \Comment{Start backward propagation$^{*3a}$}
        \State $optimizer.step()$ \Comment{Parameter update$^{*3b}$}
    \EndWhile
\EndFor
\State $training\_time\_cost \gets \text{getCurrentTime()} - start$ \Comment{Compute elapsed time}
\State \textbf{return} $training\_time\_cost$ 
\end{algorithmic}
\begin{flushleft}
\footnotesize{$^{*1-3}$: Subactivities in the training process -- forward propagation includes loss calculation and backward propagation includes parameter update.}
\end{flushleft}
\end{algorithm}

\begin{algorithm}[t]
\caption{Measuring Inference Time Cost in PyTorch Bindings}\label{alg:inference_time_tch}
\begin{algorithmic}[1]
\State $model \gets \text{loadSavedModel()}$ \Comment{Load trained model}
\State $test\_set \gets \text{loadDataset()}$ \Comment{Load pre-processed test set}
\State $start \gets \text{getCurrentTime()}$ \Comment{Start the timer}
\While{not \text{isEndOfDataset($test\_set$)}}
    \State $inputs \gets \text{getNextBatch($test\_set$)}$ \Comment{Batch data loading$^{*1}$}
    \State $preds \gets model(inputs)$ \Comment{Inference forward propagation$^{*2}$}
\EndWhile
\State $inference\_time\_cost \gets \text{getCurrentTime()} - start$ \Comment{Compute elapsed time}
\State \textbf{return} $inference\_time\_cost$
\end{algorithmic}
\begin{flushleft}
\footnotesize{$^{*1-2}$: Subactivities in the inference process.}
\end{flushleft}
\end{algorithm}

\begin{algorithm}[t]
\caption{Measuring Time Cost of a Training/Inference Subactivity in PyTorch Bindings}\label{alg:sync_time_tch}
\begin{algorithmic}[1]
\State $start \gets \text{getCurrentTime()}$ \Comment{Start the timer}
\State runSubactivity() \Comment{Execute a subactivity of training/inference}
\State $cuda.synchronize()$ \Comment{Wait for the subactivity to finish}
\State $time\_cost \gets \text{getCurrentTime()} - start$ \Comment{Compute elapsed time}
\State \textbf{return} $time\_cost$
\end{algorithmic}
\end{algorithm}

\textbf{Step 3 -- Perform inference using the trained models and record the inference correctness:}
For each model inference experiment, each binding loads a pre-trained model via the supported model loading approach(es)~(as shown in Table~\ref{tab:supported_features}) and performs inference on the test set on both CPU and GPU. In addition, bindings for the same ML framework perform inference for the same pre-trained model. We select the pre-trained models~(which are saved in Step 2) from TensorFlow and PyTorch's default Python bindings since the default bindings tend to have the best support and maintenance~\cite{mlbindings_2022}. 

\runningexample In TensorFlow's Rust binding, we load the pre-trained LeNet-1 model from TensorFlow's default Python binding via serialization to perform model inference on the test set and record the cross-binding test accuracy.

\textbf{Step 4 -- Measure and record the training time cost:}
Our primary focus is on measuring the time cost of the entire training process on GPU and recording it, as shown in Procedures \ref{alg:training_time_tch} and \ref{alg:time_cost_tf}. Due to the asynchronous nature of GPU computations~(as explained in Section~\ref{sec:background}), we only keep the code directly related to the training process in this step to ensure accurate time measurements, excluding activities like calculating correctness metrics in each epoch~(which is included in Steps 1 and 2). We also do not include the time cost of initialization processes, such as model initialization, optimizer initialization, and initial dataset loading.

Procedure~\ref{alg:training_time_tch} within PyTorch showcases its granular control over the training process. It initiates by setting up the model and optimizer, loading the training dataset, and iterating through the epochs for optimizing the model weights. For each epoch, the process starts with loading a batch of the data. Following this, forward propagation is performed to produce outputs which are used for calculating the loss values. Lastly, backward propagation is executed to calculate the gradients which guide the optimizer for updating the model parameters. In contrast, as demonstrated in Procedure~\ref{alg:time_cost_tf}, TensorFlow offers less granularity since it encapsulates the entire training process~(i.e., batch data loading, forward propagation, and backward propagation) within a single function to optimize performance.

As shown in Procedure~\ref{alg:sync_time_tch}, the granularity control in PyTorch is particularly helpful in measuring time costs for specific subactivities using the ``cuda.synchronize()" function to facilitate synchronization between the CPU and GPU. The ``cuda.synchronize()" function is only available in the Python and Rust bindings. Procedure~\ref{alg:sync_time_tch} starts a timer, runs a subactivity~(e.g., forward propagation), waits for the subactivity to finish using "cuda.synchronize()", and then computes the elapsed time.

\runningexample We train the LeNet-1 model with PyTorch’s Python binding and employ Procedure~\ref{alg:training_time_tch} to record the training time cost. In addition, we rerun the training experiment utilizing Procedure~\ref{alg:training_time_tch} with additional synchronization steps as described in Procedure~\ref{alg:sync_time_tch} to capture accurate time costs for individual subactivities.

\begin{algorithm}[t]
\caption{Measuring Training/Inference Time Cost in TensorFlow (TF) Bindings}\label{alg:time_cost_tf}
\begin{algorithmic}[1]
\State $model \gets \text{initModelAndCompile(}optimizer, loss\_function\text{)}$ \Comment{Model initialization}
\State $train\_set, test\_set \gets \text{loadDataset()}$ \Comment{Load pre-processed data}
\State $start \gets \text{getCurrentTime()}$ \Comment{Start the timer}
\State model.fit($train\_set$, $epochs$)/model.predict($test\_set$) \Comment{TF's single training/inference function}
\State $time\_cost \gets \text{getCurrentTime()} - start$ \Comment{Compute elapsed time}
\State \textbf{return} $time\_cost$
\end{algorithmic}
\end{algorithm}

\textbf{Step 5 -- Measure and record the inference time cost:}
Similar to Step 4, we measure and record the time cost of the entire inference process on both CPU and GPU following Procedures \ref{alg:inference_time_tch} and \ref{alg:time_cost_tf}. For measuring the time costs of inference subactivities~(i.e., batch data loading and forward propagation), we rerun the inference experiments employing Procedure~\ref{alg:sync_time_tch}, but only for PyTorch's Python and Rust bindings on GPU.

\runningexample In PyTorch's Python binding, we use Procedure~\ref{alg:inference_time_tch} to determine the inference time cost for the pre-trained LeNet-1 model. Furthermore, we rerun the inference experiment with additional steps from Procedure~\ref{alg:sync_time_tch} to separately record time costs for batch data loading and forward propagation.

\subsection{Supported features in studied bindings}\label{sec:binding_features}

Table~\ref{tab:supported_features} outlines the supported features by each studied binding:

\begin{itemize}
    \item \textbf{Training support:} A lack of training support in certain bindings means developers might have to use another programming language. This can be inconvenient and result in additional overhead, especially if developers are unfamiliar with the alternative language. 
    \item \textbf{Model interface support:} When certain model types are not supported in a binding, developers might still need to switch to another language to train their models.
    \item \textbf{Model loading approaches:} Loading models via serialization provides flexibility as developers don't need to define the model structure. In contrast, loading models via parameters requires the model's structure to be pre-defined. This can lead to challenges, especially when developers try to use pre-trained models.
\end{itemize}

For our training experiments in Section~\ref{sec:exp_setup}, certain bindings are exempt due to their limitations: TensorFlow's Rust and PyTorch's JavaScript bindings (which don't support training), TensorFlow's C\# binding for RNNs, and all bindings for BERT. We acknowledged the recent inclusion of support for RNNs in TensorFlow's C\# binding~(aligned with TensorFlow v2.10).\footnote{\url{https://github.com/SciSharp/TensorFlow.NET/issues/640}} However, to maintain consistency in our experimental framework, we focused on TensorFlow version 2.5.0 which is the most commonly supported version of TensorFlow by the studied bindings.

For the inference experiments, all bindings are utilized in our work, with the exception of RNNs in TensorFlow's C\# and BERT in C\# bindings for both ML frameworks. The reason is that the C\# bindings can only load models using parameters and lacks support for RNN and BERT interfaces. Unlike PyTorch's JavaScript binding which despite not supporting CNNs, RNNs, and BERT, does offer loading via serialization without the need for defining model structures.

\section{Correctness Evaluation}\label{sec:correctness_eval}

\motivation
Developers can use a binding for an ML framework in their preferred programming language to train a DL model. 
We want to observe if the DL models trained using a binding for a given ML framework have the same training accuracy as the DL models trained using the ML framework's default Python binding~(RQ1). These results can help developers understand if using a binding will achieve the same model accuracy during training and provide the same model performance for the final trained models.

In addition, it is important to ascertain if performing inference for these trained models using different bindings for a given framework will impact the accuracy. Pre-trained models have been widely used by the ML community~\cite{wolf_transformers_2020, han_pretrained_2021} and bindings can help developers to run inference with pre-trained models in different programming languages. Importantly, in high-stakes domains such as medical diagnosis and autonomous driving, accuracy is particularly important when decisions are made by ML systems~\cite{nussberger2022public}. Even a slight drop in accuracy can trigger erroneous decisions with serious implications. Hence, it is vital that bindings have the capability to achieve the same accuracy for pre-trained models as with the binding they were trained with. In RQ2, we investigate the cross-binding test accuracy of pre-trained models using the bindings for TensorFlow and PyTorch to understand whether the pre-trained models perform as we would expect them to.

Together, the bindings' impact on training correctness and inference correctness will enable us to understand the impact on the correctness of the ML software quality. 

\subsection*{RQ1: \rqone}

\begin{table}[t]
\centering	
\small
\caption{Mean/Max DTW distances of training accuracy curves for bindings in training models with the same random seed. (Highlighted numbers indicate negligible DTW distance. Py: Python; JS: JavaScript; Rs: Rust)}
\label{tab:dynamic_time_wrap}	
\begin{tabular}{lrrrrrrr}
\toprule
\multicolumn{1}{c}{\multirow{2}{*}{Model}} & \multicolumn{3}{c}{TensorFlow (mean/max DTW distance)}                                                                &                      & \multicolumn{3}{c}{PyTorch (mean/max DTW distance)}                                                                         \\ \cmidrule{2-4} \cmidrule{6-8} 
\multicolumn{1}{c}{}                       & \multicolumn{1}{c}{Py-C\#}      & \multicolumn{1}{c}{Py-JS} & \multicolumn{1}{c}{JS-C\#}      & \multicolumn{1}{c}{} & \multicolumn{1}{c}{Py-C\#}      & \multicolumn{1}{c}{Py-Rs}   & \multicolumn{1}{c}{Rs-C\#}      \\ \midrule
LeNet-1                                    & \badcolor{0.005/0.006}                           & \okcolor{0.000/0.000}                     & \badcolor{0.005/0.006}                           &                      & \okcolor{0.000/0.000}                           & \okcolor{0.000/0.000}                           & \okcolor{0.000/0.000}                           \\
LeNet-5                                    & \badcolor{0.003/0.004}                           & \okcolor{0.000/0.000}                     & \badcolor{0.003/0.004}                           &                      & \okcolor{0.000/0.000}                           & \okcolor{0.000/0.000}                           & \okcolor{0.000/0.000}                           \\
VGG-16                                     & \badcolor{0.018/0.019} & \badcolor{0.005/0.006}                     & \badcolor{0.018/0.019} &                      & \badcolor{0.007/0.010}                           & \badcolor{0.002/0.003}                           & \badcolor{0.008/0.010}                           \\
LSTM                                       & -                               & \badcolor{0.008/0.012}                     & -                               &                      & \badcolor{0.008/0.009}                           & \badcolor{0.009/0.011} & \badcolor{0.010/0.011}                           \\
GRU                                        & -                               & \badcolor{0.010/0.012}                     & -                               &                      & \badcolor{0.010/0.011} & \badcolor{0.008/0.009}                           & \badcolor{0.009/0.010} \\ \bottomrule
\end{tabular}
\end{table}

\begin{figure*}[t]
	\centering
	\includegraphics[width=\textwidth]{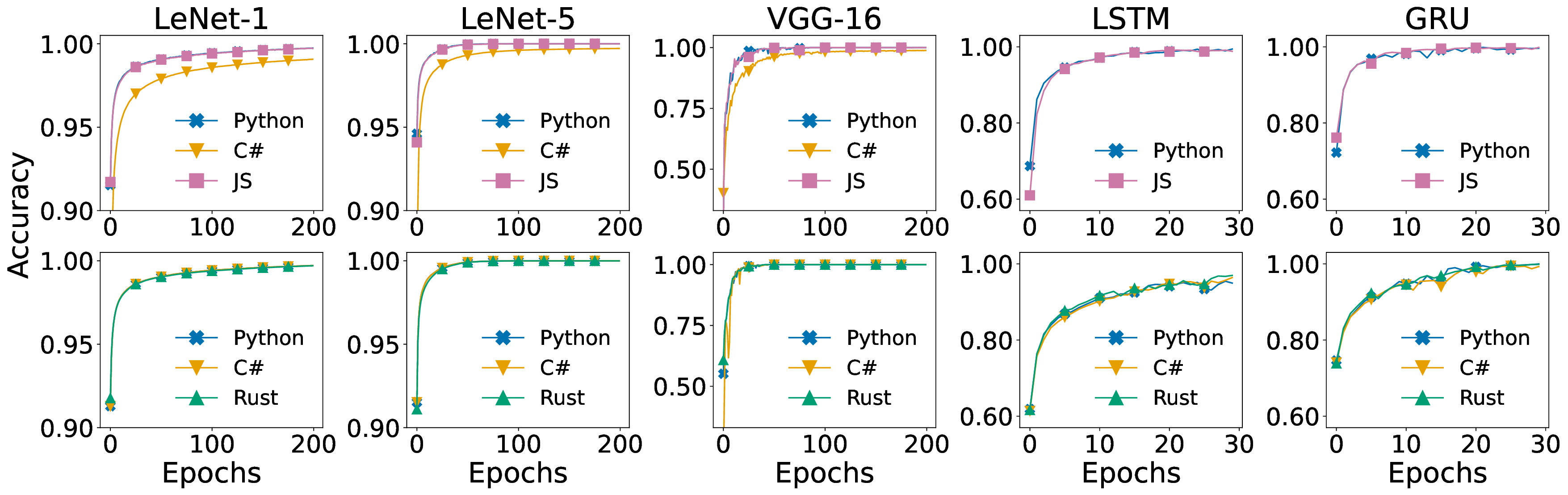}
	\caption{Mean training accuracy curves of LeNet-1, LeNet-5, VGG-16, LSTM, and GRU on GPU in bindings for TensorFlow~(first row) and PyTorch~(second row).}
	\label{fig:development_training_errors}
\end{figure*}

\approach
We employ both dynamic time warping~(DTW)~\cite{salvador_fastdtw_2007} for analyzing training accuracy curves and the Mann-Whitney U test~\cite{Mann1947OnAT} for comparing the performance metrics of the final trained models. We chose DTW due to its ability to analyze time-series data, which allows us to investigate whether different bindings follow the same trajectory during training. DTW calculates the distance between the training accuracy curves of the bindings~(e.g., between TensorFlow's Python and C\# binding) for training the same model. DTW is widely used as a distance measurement for time series data since it can manage time distortion by aligning two time series before computing the distance, which is more accurate than the Euclidean distance~\cite{ding_dtw_2008}. We normalize the calculated DTW distances between 0 to 1 to interpret the results. A normalized DTW distance of 0 means that the difference between the two curves is negligible. 

In addition, we calculate the test accuracy, F1-score, and AUC-ROC for the final trained models to compare their classification performance. For each metric, we perform the Mann-Whitney U test~\cite{Mann1947OnAT} separately at a significance level of $\alpha=0.05$ to determine if the values obtained from different bindings are significantly different. We computed Cliff’s delta~$d$~\cite{Cliff} effect size to quantify the difference based on the following thresholds~\cite{Cliff_threshold}:

\begin{equation} \label{effectsize}
\mathrm{Effect \ size} = 
\left\{
\begin{array}{ll}
	negligible,  & \mathrm{if} \ |d|  \le 0.147 \\
	small,  & \mathrm{if} \ 0.147 < |d|  \le 0.33 \\
	medium,  & \mathrm{if} \ 0.33 < |d|  \le 0.474 \\
	large,  & \mathrm{if} \ 0.474 < |d|  \le 1 \\
\end{array}\right.
\end{equation}

\findings
\emp{Bindings can have different training accuracy curves when training DL models under the same configuration (i.e., model structure, training data, hyperparameters, and random seed).} Table~\ref{tab:dynamic_time_wrap} reports the mean and maximum DTW distances for the training curves between bindings. Moreover, Figure~\ref{fig:development_training_errors} presents the mean training accuracy curves of the models~(out of the five training processes) that have the best test accuracy after the last epoch. The figure and table show that bindings can have quite different training accuracy curves according to the DTW distance when using the same training configuration. For example, the distances between the curves of TensorFlow's C\# binding and the other two bindings are relatively large for LeNet-1, LeNet-5, and VGG-16 models. Another example is that all PyTorch bindings have a relatively large distance between the curves for the RNN models compared to the distances in the CNN models. One reason could be the differential numerical precision across programming languages. For example, Python supports arbitrary-precision arithmetic, while languages like Rust and C\# typically operate with fixed precision. These variations in numerical precision might spawn minor differences in mathematical computation outputs. These minor differences might accumulate over numerous iterations during model training, resulting in variations in the final model accuracy. In contrast, bindings can exhibit nearly the same behaviour for training some DL models; the training accuracy curves of the LeNet models differ negligibly between TensorFlow's Python and JavaScript bindings, as well as between PyTorch's bindings.

\begin{table}[t]
\small
    \centering
    \caption{The average test accuracy (Acc), F1-score (F1), and AUC-ROC (AUC) for TensorFlow and PyTorch bindings. (Statistically significant differences between bindings are highlighted in bold. Py: Python; JS: JavaScript; Rs: Rust; MD: Max Diff; ES: Effect Size)}
    \label{tab:test_acc}
\begin{tabular}{llccccc||lccccc}
\toprule
                     & \multicolumn{6}{c||}{TensorFlow}                                                                                    & \multicolumn{6}{c}{PyTorch}                                                                                 \\
                     &                                                    & LN1        & LN5  & VGG         & LSTM        & GRU         &                                                    & LN1  & LN5        & VGG         & LSTM & GRU         \\ \midrule
\multirow{7}{*}{Acc} & Py                                                 & 98.8        & 98.9  & 84.8        & 83.7        & 85.0        & Py                                                 & 98.8  & 98.9        & 86.2        & 86.5 & 87.9        \\
                     & C\#                                                & 98.6        & 98.9  & 83.8        &             &             & C\#                                                & 98.8  & 99.0        & 86.2        & 87.3 & 85.5        \\
                     & JS                                                 & 98.8        & 99.0  & 85.6        & 84.2        & 84.7        & Rs                                                 & 98.8  & 98.9        & 85.6        & 87.4 & 87.0        \\ \cline{2-13}
                     & MD & \textbf{0.2} & 0.1   & \textbf{1.9} & 0.6         & 0.3         
                     & MD & 0.0   & 0.1         & \textbf{0.6} & 0.8  & \textbf{2.5} \\ 
                     & $p$         & \textbf{0.01}        & 0.40  & \textbf{0.01}        & 0.10        & 0.15
                     & $p$         & 0.68        & 0.31  & \textbf{0.03}        & 0.10        & \textbf{0.01}        \\
                     & ES          & \textbf{large}        & -  & \textbf{large}        & -        & -       
                     & ES          & -        & -  & \textbf{large}        & -        & \textbf{large}       \\
                     \midrule
\multirow{5}{*}{F1}  & Py                                                 & 98.8        & 98.9  & 84.7        & 83.5        & 85.0        & Py                                                 & 98.8  & 99.0        & 86.3        & 86.7 & 87.9        \\
                     & C\#                                                & 98.6        & 98.9  & 83.8        &             &             & C\#                                                & 98.8  & 99.0        & 86.1        & 87.2 & 85.1        \\
                     & JS                                                 & 98.8        & 99.0  & 85.6        & 83.8        & 84.7        & Rs                                                 & 98.9  & 98.9        & 85.6        & 87.2 & 86.9        \\ \cline{2-13}
                     & MD & \textbf{0.2} & 0.1   & \textbf{1.9} & 0.3         & 0.3                 
                     & MD & 0.1   & \textbf{0.1} & \textbf{0.7} & 0.5  & \textbf{2.8} \\
                     & $p$         & \textbf{0.01}        & 0.42  & \textbf{0.01}        & 0.22        & 0.15
                     & $p$         & 0.06        & \textbf{0.01}  & \textbf{0.01}        & 0.10        & \textbf{0.01}        \\
                     & ES          & \textbf{large}        & -  & \textbf{large}        & -        & -       
                     & ES          & -        & \textbf{large}  & \textbf{large}        & -        & \textbf{large}       \\
                     \midrule
\multirow{5}{*}{AUC} & Py                                                 & 100.0       & 100.0 & 98.2        & 91.7        & 92.3        & Py                                                 & 100.0 & 100.0       & 98.5        & 94.1 & 94.3        \\
                     & C\#                                                & 100.0       & 100.0 & 97.3        &             &             & C\#                                                & 100.0 & 100.0       & 98.5        & 94.6 & 92.9        \\
                     & JS                                                 & 100.0       & 100.0 & 98.4        & 92.3        & 91.9        & Rs                                                 & 100.0 & 100.0       & 98.3        & 94.5 & 93.8        \\ \cline{2-13}
                     & MD & 0.0         & 0.0   & \textbf{1.1} & \textbf{0.6} & \textbf{0.5}         
                     & MD & 0.0   & 0.0         & \textbf{0.2} & 0.5  & \textbf{1.5} \\
                     & $p$         & 0.10        & 0.84  & \textbf{0.01}        & \textbf{0.01}        & \textbf{0.01}
                     & $p$         & 0.55        & 0.42  & \textbf{0.01}        & 0.10        & \textbf{0.01}        \\
                     & ES          & -        & -  & \textbf{large}        & \textbf{large}        & \textbf{large}       
                     & ES          & -        & -  & \textbf{large}        & -        & \textbf{large}       \\
                     \bottomrule
\end{tabular}
\end{table}

\emp{The trained models produced by certain bindings can perform worse than the models produced by other bindings for the same ML framework.} Table~\ref{tab:test_acc} shows the test accuracy, F1-score, and AUC-ROC for the trained models produced by bindings can be different. For the trained VGG-16 models, the Mann-Whitney U test reveals significant differences between bindings for both frameworks in these metrics with large effect sizes. This pattern is also observed in the trained GRU models in PyTorch's bindings. Specifically, while the test accuracy and F1-score of the trained LeNet-1 models have statistically significant differences between bindings for TensorFlow, the AUC-ROC values of LeNet models in TensorFlow and PyTorch bindings are close (all rounded up to 100 in Table~\ref{tab:test_acc}). Furthermore, we observed some models produced by non-Python bindings have higher values of the metrics than the models produced by the default Python bindings, e.g., the VGG-16 model produced by TensorFlow's JavaScript binding.


\begin{Summary}{Summary of RQ1}{}
TensorFlow and PyTorch bindings can have different training accuracy curves for training the same DL models even when using the same configuration. In addition, the test accuracy of the final trained models can be slightly different. Hence, developers should not assume that all bindings offer the same level of correctness and should verify the model's correctness when utilizing a binding for training.
\end{Summary}

\subsection*{RQ2: \rqtwo}

\begin{figure}[t]
	\centering
	\includegraphics[width=.7\columnwidth]{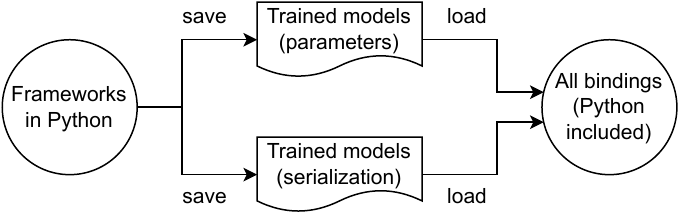}
	\caption{All bindings load the trained models that are saved by the default Python bindings for ML frameworks.}
	\label{fig:rq2_approach}
\end{figure}

\approach
We conducted inference experiments with all bindings using pre-trained models produced by the default Python bindings for TensorFlow and PyTorch (see Figure~\ref{fig:rq2_approach}). We loaded the pre-trained models using the supported loading approach(es) and recorded the cross-binding test accuracy on both CPU and GPU for each binding. If the cross-binding test accuracy of a pre-trained model in a binding shows a 0\% difference compared to the test accuracy when the model was initially trained, we considered the test accuracy \enquote{reproduced} by that binding. Any non-zero difference resulted in a \enquote{failed} mark. Since some bindings only support one way of loading models~(as shown in Table~\ref{tab:supported_features}), we marked the result as \enquote{unsupported} if the loading approach is not supported by a binding. 

\begin{figure}[t]
	\centering
	\begin{subfigure}[b]{0.55\columnwidth}
		\centering
		\includegraphics[width=\textwidth]{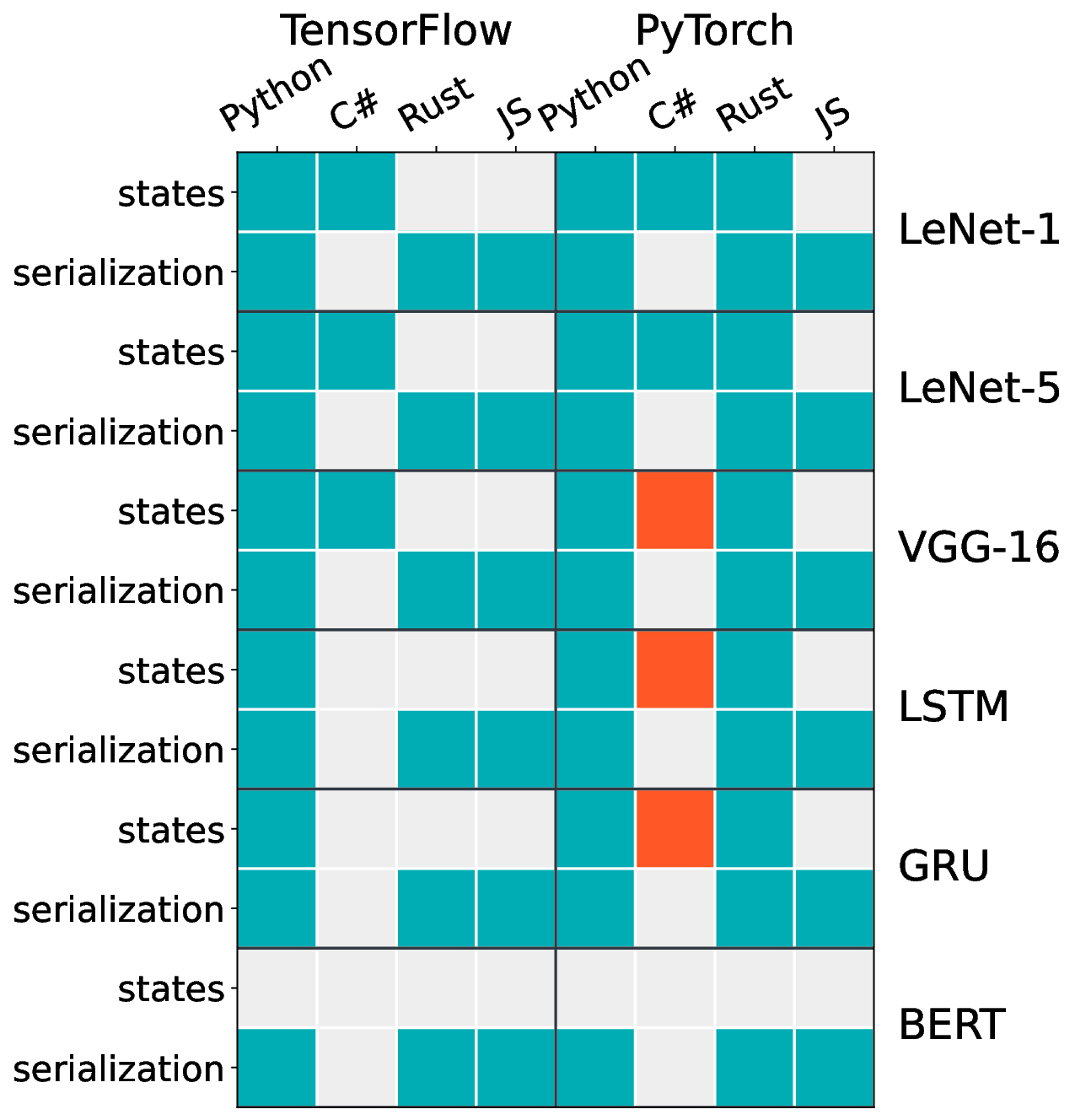}
	\end{subfigure}
	\hfill
	\begin{subfigure}[h]{0.45\columnwidth}
		\centering
		\includegraphics[width=\textwidth]{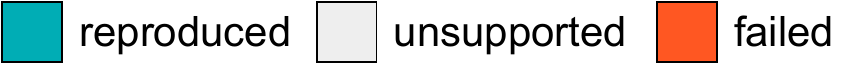}
	\end{subfigure}
	\caption{Results of reproducing the test accuracy of pre-trained models in TensorFlow and PyTorch bindings on the CPU and GPU (the results are identical). Note: the failed cases in the PyTorch's C\# binding were fixed in a newer version of the binding.}
	\label{fig:reproduce}
\end{figure}

\findings
\emp{The test accuracy of pre-trained models can be reproduced across bindings in different languages for the same ML framework.} Figure~\ref{fig:reproduce} shows that only PyTorch's C\# binding failed to reproduce the test accuracy in the saved VGG-16, LSTM, and GRU models. We noticed that the differences in the test accuracy in these three models are all within 1\% and the root cause of the reproduction failure is a bug that results in ``eval() and train() methods not being properly propagated to all submodules''.\footnote{See \url{https://github.com/dotnet/TorchSharp/pull/501} and \url{https://github.com/dotnet/TorchSharp/issues/500}} This bug prevents setting the model to evaluation mode, hence, the dropout layers of these three models are not disabled which leads to different cross-binding test accuracy. This bug is fixed in version \verfmt{0.96.0} which does not support PyTorch~\verfmt{1.9.0} but targets version \verfmt{1.10.0}. In other words, the saved models can be reproduced in the newer version of PyTorch's C\# binding. For consistency, we still use the \verfmt{0.93.9} version of this binding for the other experiments.

\emp{Bindings can reproduce the test accuracy of pre-trained models via different loading approaches and on different types of processing units~(i.e., CPU and GPU).} As shown in Figure~\ref{fig:reproduce}, PyTorch's Python and Rust bindings and TensorFlow's Python binding support both loading via parameters and serialization, and both loading approaches can reproduce the test accuracy of the pre-trained models. In addition, we noticed that bindings can reproduce the test accuracy of pre-trained models on both CPU and GPU.

\begin{Summary}{Summary of RQ2}{}
TensorFlow and PyTorch bindings can perform inference using pre-trained models and reproduce the same test accuracy as when the models were originally trained. This correctness property holds true whether model inference is performed on CPU or GPU. As a result, developers can leverage the capabilities of pre-trained models while still being able to use the model in their preferred language. 
\end{Summary}

\section{Time Cost Evaluation}\label{sec:cost_eval}
\motivation
In RQ1 and RQ2, we studied the impact of bindings for ML frameworks on correctness, however, the impact of bindings on time cost remains unknown. 
Given the time-consuming nature of model training and model inference for ML frameworks, it is important to investigate how a binding may impact the time cost. Studies show that runtime efficiency and energy consumption can vary across programming languages~\cite{prechelt_pls_2000, nanz_pls_2015, pereira_pls_2017}. Consequently, these differences may have an impact on the time cost of training and inference when using different bindings.

Thus, in RQ3, we study the time cost of training DL models with bindings in order to offer developers more information about the overhead or advantage in terms of time cost when training with a binding. In RQ4, we study the inference time of pre-trained models in bindings. The time of utilizing bindings in model inference can be a crucial consideration for developers since model inference typically takes place~(as a part of the product) in the production environment, which may have limited resources. The findings can help developers decide whether or not to utilize a binding for model inference in their project.

\subsection*{RQ3: \rqthree}

\begin{figure*}[t]
	\centering
	\includegraphics[width=\textwidth]{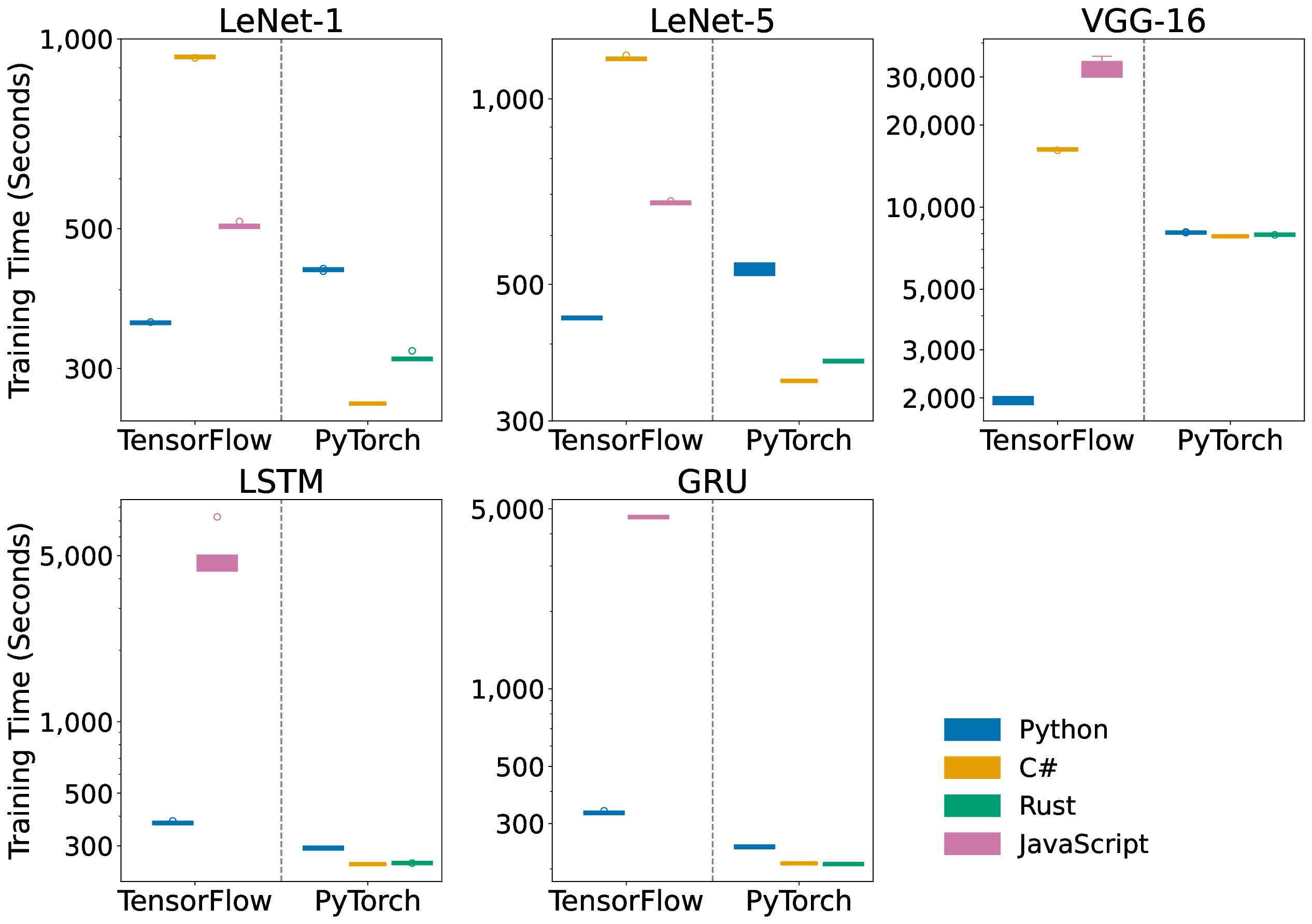}
	\caption{Training time distributions when training models in TensorFlow and PyTorch bindings on the GPU.}
	\label{fig:training_time_costs}
\end{figure*}

\approach
To study the difference in training time across bindings, we performed the Mann-Whitney U test~\cite{Mann1947OnAT} using the Bonferroni correction~\cite{shaffer_multiple_1995} to adjust the significance level for multiple comparisons. Specifically, for an initial significance level of $\alpha=0.05$, we adjusted the significance level to $\frac{\alpha}{n}$~(where $n$ is the number of comparisons made) to determine whether the distributions of the training times of the default Python bindings and the non-Python bindings, which trained the same model for the same framework, are significantly different. For example, the LeNet-1 model in TensorFlow bindings, we performed Bonferroni-corrected Mann-Whitney U test between the Python and C\# bindings and Python and JavaScript bindings with an adjusted significance level of $\frac{\alpha}{2}=0.025$. We also computed Cliff's delta~$d$~\cite{Cliff} effect size to quantify the difference based on Equation~\ref{effectsize} in Section~\ref{sec:correctness_eval}.



\findings
\emp{Training times can differ greatly across bindings for the same ML framework.} Figure~\ref{fig:training_time_costs} shows the training time distributions on GPU for the studied models across the studied bindings. The Bonferroni-corrected Mann-Whitney U test shows that the training time distributions of the same model are all significantly different between the default Python bindings and the other bindings for the same framework and the effect sizes are all large. In addition, the difference in training time of bindings for the same ML framework can be very large when training certain models. For example, the median training time of TensorFlow's JavaScript binding for the VGG-16 model is 15 times larger than its Python binding~(32,783 vs. 1,991 seconds).


\emp{PyTorch's default Python binding has the slowest training time for the studied models.} Figure~\ref{fig:training_time_costs} shows that PyTorch's Python binding is more than two times slower than the other two bindings for training LeNet models. However, we note that the training time difference between PyTorch's Python binding and other bindings for the VGG-16, LSTM, and GRU models is relatively small~(less than 15\%). In contrast, TensorFlow's default Python binding has the fastest training time in the studied models.

\begin{table}[t]
\centering
\caption{Time costs (in seconds) of the subactivities in the training process using PyTorch's Python and Rust bindings on GPU.}
\label{tab:tch_sync_training_costs}
\begin{tabular}{llrrrr}
\toprule
                         &        & Load batch data & Forward & Backward & Total  \\ \midrule
\multirow{2}{*}{LeNet-1} & Python & 148.9     & 76.2    & 240.2    & 465.7  \\
                         & Rust   & 23.5      & 69.7    & 239.2    & 332.5  \\ \midrule
\multirow{2}{*}{LeNet-5} & Python & 167.4     & 114.8   & 293.2    & 576.4  \\
                         & Rust   & 24.1      & 94.3    & 278.8    & 397.4  \\ \midrule
\multirow{2}{*}{VGG-16}  & Python & 89.2      & 7094.0  & 1557.8   & 8741.2 \\
                         & Rust   & 31.9      & 6470.4  & 1469.5   & 7971.8 \\ \midrule
\multirow{2}{*}{LSTM}    & Python & 8.2       & 95.4    & 188.8    & 292.2  \\
                         & Rust   & 0.6       & 83.5    & 165.3    & 249.4  \\ \midrule
\multirow{2}{*}{GRU}     & Python & 8.5       & 84.0    & 151.0    & 242.8  \\
                         & Rust   & 0.6       & 72.5    & 130.3    & 203.5  \\ \bottomrule
\end{tabular}
\end{table}

\emp{Batch data loading time affects the training cost of PyTorch's Python binding.} As shown in Table~\ref{tab:tch_sync_training_costs}, PyTorch's Python binding has a long batch data loading time, which is notably slower (between 4 to 14 times) than the Rust binding for all studied models. Specifically, For LeNet models, the Python binding's batch data loading times account for roughly 30\% of the training cost, whereas the Rust binding's batch data loading for the same models consumes less than 10\% of the training cost. Furthermore, the Python binding consistently underperforms the Rust binding during both forward and backward propagation phases in the studied models.

The observed variations in batch data loading times between bindings suggest that the native speed of a programming language~\cite{prechelt_pls_2000, nanz_pls_2015, pereira_pls_2017} is an important factor that influences the performance of a binding. However, there could be other factors involved in the implementation of bindings. For example, these factors could include overheads arising from differences in data structure implementations and initialization routines. Additionally, the overhead of the marshalling mechanism~\cite{ekblad_ffi_2015, jeremy_ffi_2018, bruni_ffi_2013} implemented to convert data between the binding’s programming language and the ML framework could impact efficiency. Finally, the way the binding interacts with the ML framework's lower-level APIs, such as those for memory management and tensor operations, could also play a crucial role in performance differences.


\begin{Summary}{Summary of RQ3}{}
Training times for training the same DL models differ significantly between the default Python bindings and the non-Python bindings for the same ML framework. Surprisingly, non-Python bindings for PyTorch are even faster in training the studied models than the default Python binding. Hence, choosing the right binding can help developers to lower the training time cost for certain models.
\end{Summary}

\subsection*{RQ4: \rqfour}

\begin{figure*}[t]
	\centering
	\includegraphics[width=\textwidth]{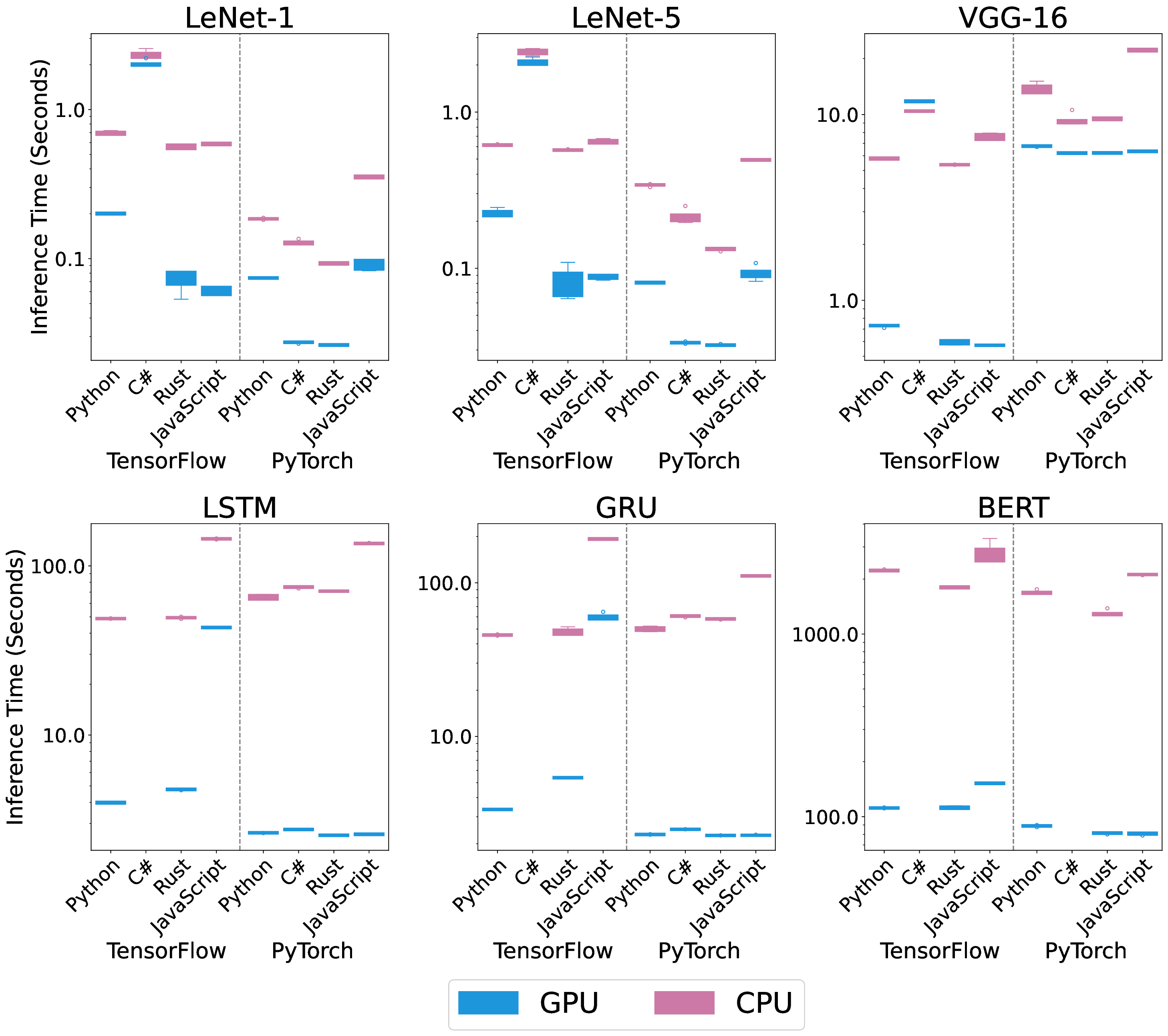}
	\caption{Inference time distributions for pre-trained models in TensorFlow~(TF) and PyTorch~(PT) bindings on the CPU and GPU.}
	\label{fig:deployment_time_costs}
\end{figure*}

\approach
We followed the same process as shown in Figure~\ref{fig:rq2_approach} and investigated the inference time of each model on both CPU and GPU. We performed the Bonferroni-corrected Mann-Whitney U test on the recorded inference time distributions between the default Python bindings and the non-Python bindings, grouped by the same framework, model, and processing unit~(CPU or GPU). We also computed Cliff's Delta effect size as described in RQ3. 

\findings
\emp{The inference time of the same pre-trained model differs greatly between the default Python bindings and the other bindings for the same ML framework.} Figure~\ref{fig:deployment_time_costs} shows the distributions of the inference time of the pre-trained models in the studied bindings. The results of the Bonferroni-corrected Mann-Whitney U test and Cliff's Delta~$d$ show that the Python and non-Python bindings for the same ML framework have significantly different inference times for the same model on the same processing unit (i.e., CPU and GPU) and the effect size is large, except for the TensorFlow bindings for LSTM on CPU and for BERT on GPU where the Python binding has similar inference time costs as the Rust binding. We observed that the default Python bindings for TensorFlow and PyTorch do not always offer the best inference time for all studied pre-trained models, with Rust bindings often outperforming them. On the other hand, TensorFlow's C\# binding has the worst performance for the studied models on both CPU and GPU, and PyTorch's JavaScript binding has the worst performance on CPU. Moreover, the performance gap in model inference time can be very large, for example, TensorFlow's Python binding is 17 times as fast as the JavaScript binding for the GRU model on the GPU~(3.35 vs. 58.32 seconds).


\begin{table}[t]
\centering
\caption{Time costs (in seconds) of the subactivities in the inference process using PyTorch's Python and Rust bindings on GPU.}
\label{tab:tch_sync_inference_costs}
\begin{tabular}{llrrr}
\toprule
                         &        & Load batch data & Forward & Total \\ \midrule
\multirow{2}{*}{LeNet-1} & Python & 0.06      & 0.02    & 0.08  \\
                         & Rust   & 0.01      & 0.02    & 0.03  \\ \midrule
\multirow{2}{*}{LeNet-5} & Python & 0.06      & 0.03    & 0.08  \\
                         & Rust   & 0.01      & 0.02    & 0.03  \\ \midrule
\multirow{2}{*}{VGG-16}  & Python & 0.12      & 6.74    & 6.86  \\
                         & Rust   & 0.03      & 6.26    & 6.29  \\ \midrule
\multirow{2}{*}{LSTM}    & Python & 0.15      & 2.61    & 2.76  \\
                         & Rust   & 0.01      & 2.57    & 2.58  \\ \midrule
\multirow{2}{*}{GRU}     & Python & 0.15      & 2.27    & 2.41  \\
                         & Rust   & 0.03      & 2.25    & 2.28  \\ \midrule
\multirow{2}{*}{BERT}    & Python & 0.13      & 88.86   & 88.99 \\
                         & Rust   & 0.12      & 81.82   & 81.94 \\ \bottomrule
\end{tabular}
\end{table}

\emp{Inference time differences in PyTorch arise from both batch data loading and forward propagation speed.} Table~\ref{tab:tch_sync_inference_costs} shows that the majority of the inference cost is allocated towards forward propagation and the Rust binding outperforms the Python binding in this regard. As we observed the same pattern in RQ3, the Rust binding also demonstrates faster batch data loading times compared to the Python binding across all studied models. Although both bindings leverage PyTorch's computational core, which is written in C/C++ and predominantly runs computations on GPUs, the variations in time costs can be attributed to overheads introduced by the bindings themselves. 

\emp{Certain bindings on the CPU may have a faster inference time than other bindings on the GPU for the same pre-trained model.} Generally, inference time for pre-trained models on GPU outperforms CPU in bindings for both studied frameworks~(as shown in Figure~\ref{fig:deployment_time_costs}). However, we found that for the same framework, one binding that runs inference on CPU can outperform another binding that runs on GPU for the same pre-trained model. For example, the Rust binding for TensorFlow is faster on CPU than the C\# binding on GPU for LeNet and VGG-16 models, as well as faster on CPU than the JavaScript binding on GPU for GRU model. Furthermore, we noticed that TensorFlow's C\# binding in model inference on CPU is similar to or even faster than on GPU. According to the maintainer of the C\# binding, the reason could be that \enquote{there is I/O cost underlying}\footnote{\url{https://github.com/SciSharp/TensorFlow.NET/issues/876}} model inference on GPU. 

\emp{Certain bindings lack support for certain features which leads to a slower inference time.} We noticed that TensorFlow's JavaScript binding cannot load a GRU model with \enquote{reset\_after=True}\footnote{\url{https://github.com/tensorflow/tfjs/issues/4621}}, either by loading parameters or through serialization. However, \enquote{reset\_after=True} is the default setting in the framework~(and other bindings) to enable the \enquote{fast cuDNN implementation}, which speeds up the inference of the GRU model\footnote{\url{https://www.tensorflow.org/api_docs/python/tf/keras/layers/GRU}} This unsupported feature can be one of the reasons behind the large increase of GRU inference time in TensorFlow's JavaScript binding~(256.5 seconds) compared to the inference time of the default Python binding~(3.6 seconds).

\begin{Summary}{Summary of RQ4}{}
TensorFlow and PyTorch bindings have various inference times for the same pre-trained models on CPU and GPU. Remarkably, the inference time of certain models in bindings on the CPU can be faster than other bindings for the same framework on GPU. Therefore, developers can experiment and choose the fastest binding for their usage scenario.
\end{Summary}

\section{Implications}\label{sec:implications}

\subsection{Implications for developers}
\emp{Developers are not limited to writing their projects in Python when using an ML framework.} Although Python dominates the development in ML~\cite{ben_braiek_open-closed_2018, nguyen_machine_2019}, developers can also use bindings in other programming languages. Our results in Section~\ref{sec:correctness_eval} shows that non-default bindings for TensorFlow and PyTorch can have the same inference accuracy of a pre-trained model as the default Python binding and sometimes even faster performance. We recommend developers use the binding in their preferred programming language for either model training or inference if supported by the binding. Hence, developers can save time and effort when adopting ML techniques in their projects without having to settle for non-mature ML frameworks that might be available in the language that their current software is programmed in. For instance, in Integration Scenario 1 of Section~\ref{sec:introduction}, Anna can use the JavaScript binding to perform inference with pre-trained models provided by the ML team.

\emp{Developers can use a binding for an ML framework which has a shorter training time for a certain model and perform inference on the trained model in another binding that has a shorter inference time based on task and requirements.} Bindings for an ML framework have various training times and inference times for ML models~(Section~\ref{sec:cost_eval}). Hence, developers can choose different bindings which are faster for a certain model in training and inference respectively since the accuracy of pre-trained models can be reproduced across bindings for the same framework~(Section~\ref{sec:correctness_eval}). We suggest that developers refer to an existing benchmark like ours or conduct experiments themselves based on our replication package~\cite{replication_package}. For example, when using TensorFlow for LeNet models as described in Integration Scenario 3 of Section~\ref{sec:introduction}, Anna can train the models using the default Python binding for TensorFlow and then run inference for the trained model in the Rust binding with the assistance of a hired expert to save time and computational resources, as this factor is critical in their project requirements.

\emp{Developers should perform a sanity check before using a model that was trained by a binding other than the default Python binding.} Bindings corresponding to different languages can have different training accuracy curves while training the same model, and the final trained model can behave differently~(as discussed in Section~\ref{sec:correctness_eval}). Since the Python bindings are the default binding for most ML frameworks, these Python bindings have a larger user base and better support than other bindings. We suggest that developers perform a sanity check on the trained model if they are using a binding other than the default Python binding before deploying the models to the production environment.

\emp{In resource-limited scenarios~(e.g., CPU only), developers may prefer or need to use a non-default binding for model inference.} Traditionally, model inference is done using a GPU due to the superior inference time of GPUs~\cite{buber_gpu_2018, li_cpu_gpu_2019}. However, GPUs are expensive and not available in all scenarios. We found that the bindings for ML frameworks can be fast for running inference on CPU for some pre-trained models~(Section~\ref{sec:cost_eval}). Developers can use such bindings if the production environment does not contain a GPU or the computational resource is limited. For example, in Integration Scenario 3 of Section~\ref{sec:introduction}, if Anna is using PyTorch for LeNet models and there is no GPU available in the production environment, she can use PyTorch's Rust binding on CPU with expert assistance. The inference time of LeNet models in the Rust binding on CPU is faster than the default Python binding both on CPU and GPU. This is particularly beneficial for constrained environments like the Internet-of-Things~(IoT) devices~(e.g., unmanned aerial vehicles) where resource availability is often limited~\cite{albanese_iot_2022, jouhari_iot_2022, s_arish_iot_2019, fedorov_iot_2019}.



\subsection{Implications for binding owners}
\emp{Binding owners should include performance benchmarks for their binding.} We found that bindings can have very different training and inference times for ML models~(Section~\ref{sec:cost_eval}), yet this information is not well documented. To address this, we suggest that binding owners introduce performance benchmarks of training and running inference for some frequently used ML models~(e.g., VGG models) and record the results in their documentation. This way, developers be aware of the trade-off between choosing a familiar language and the potential impact on time cost for various DL models. For example, the performance benchmarks can help Anna in Integration Scenario 2 of Section~\ref{sec:introduction} to make informed decisions when choosing a familiar language for training while considering the potential impact on time cost.


\subsection{Implications for researchers}
\emp{Researchers should investigate the impact of ML framework bindings on large-scale models and datasets.} Our findings provide a starting point, but further research is needed to fully understand how binding choices influence performance in large-scale models. While full-parameter fine-tuning can be computationally expensive, parameter-efficient techniques like Low-Rank Adaptation~(LoRA)~\cite{hu_lora_2022} offer a cost-effective alternative. However, LoRA's experimental status in HuggingFace\footnote{\url{https://huggingface.co/docs/diffusers/en/training/lora}} and its lack of binding support highlight a direction for further research. We suggest future research adopt our methodology~(see our replication package~\cite{replication_package}), starting with representative data subsets and smaller model variants~(e.g., the 7 billion parameter variant of Llama 2~\cite{touvron_llama2_2023}). This approach could provide valuable early insights into potential performance variations before committing to full-scale experiments.

\emp{Researchers should investigate methods to enhance the interoperability and compatibility of pre-trained models across different bindings for ML frameworks.} Our findings demonstrate that pre-trained models can be used across different bindings for the same ML framework with the same level of accuracy~(as shown in Section~\ref{sec:correctness_eval}). However, some models may not be supported or may have a slower inference time when utilizing certain bindings~(as discussed in Section~\ref{sec:cost_eval}). While developers and binding owners focus on the implementation of bindings, we suggest researchers explore ways to contribute at a higher level: by devising algorithms, methodologies, or protocols to increase the interoperability and compatibility of pre-trained models across different bindings, benefiting a diverse developer base.

\emp{Researchers should study the patterns and origins of bugs in bindings for ML frameworks.} We found that bugs in bindings for ML frameworks have an impact on the model inference correctness~(Section~\ref{sec:correctness_eval}). While the immediate resolution of bugs in bindings is an engineering concern, a deeper analysis of these issues can provide invaluable insights into software design and testing paradigms for bindings. Although researchers have previously studied bugs in ML frameworks~\cite{Chen_bugs_2022, Jia_bugs_2020, jia_bugs_2021}, there has been no research specifically on bugs in the bindings for ML frameworks or other libraries. We encourage researchers to systematically analyze the bugs in bindings and provide guidelines for maintainers to avoid introducing such bugs.



\section{Related work}\label{sec:relatedwork}


\subsection{Impact of ML frameworks on ML software correctness}
Researchers have studied the correctness of ML frameworks. However, no one has studied how bindings for those frameworks impact the correctness of the ML software that is created with them. 
The study by Guo~\etal\cite{guo_empirical_2019} is the closest related to our work. However, even though they included several bindings in their study, their work differs from ours as they focus on the impact on ML software quality of using different ML frameworks and executing ML models on different computing devices (such as PC and various types of mobile devices). In contrast, we run our experiments on the same device but we study the impact of various bindings on ML software quality. Hence, we can reason about the impact of using a binding, while in Guo~et~al.'s study, the different devices make this impossible.

Several others have focused on comparing the accuracy of the same model across ML frameworks. Chirodea~\etal\cite{chirodea_comparison_2021} compared a CNN model that was built with TensorFlow and PyTorch and found that these two frameworks have similar training curves but the final trained model has a lower accuracy in PyTorch. Gevorkyan~\etal\cite{gevorkyan_review_2019} gave an overview of five ML frameworks and compared the accuracy of training a neural network for the MNIST dataset. They reported that the final trained model has a lower accuracy in TensorFlow than in other frameworks. Moreover, Elshawi~\etal\cite{elshawi_dlbench_2021} conducted training experiments for six ML frameworks by using the default configuration and reported that certain frameworks have better performance than the other frameworks on the same model~(e.g., Chainer on the LSTM model).

\subsection{Impact of ML frameworks on ML software time cost}
Many studies have compared the time cost across ML frameworks. In a comparison of the training and inference time for a CNN architecture using PyTorch and TensorFlow, Chirodea~\etal\cite{chirodea_comparison_2021} found that PyTorch is faster in both model training and inference than TensorFlow. However, 
Gevorkyan~\etal\cite{gevorkyan_review_2019} showed that PyTorch has the worst training time for neural networks among five studied ML frameworks. In our work, we compared the training and inference time across bindings for the same ML frameworks. 

Several studies have focused on the time cost of ML frameworks on different hardware devices. Buber and Diri~\cite{buber_gpu_2018} compared the running time of DL models on CPU and GPU and found that GPU is faster. Jain~\etal\cite{jain_performance_2019} focused on the performance of training DNN models on CPU with TensorFlow and PyTorch. They show that multi-processing provides better training performance when using a single-node CPU. For mobile and embedded devices, Luo~\etal\cite{luo_comparison_2020} introduced a benchmark suite to evaluate the inference time cost based on six different neural networks.

\subsection{Impact of ML frameworks on ML software reproducibility}
Reproducibility has become a challenge in ML research~\cite{Gundersen_Kjensmo_2018, Matthew_Reproducibility_2021, Tatman_Reproducibility_2018}. Liu~\etal\cite{Liu_Reproducibility_2021} surveyed 141 published ML papers and conducted experiments for four ML models. The results showed that most studies do not provide a replication package and the models are highly sensitive to the size of test data. In addition, Isdahl and Gundersen~\cite{Isdahl_out_2019} introduced a framework to evaluate the support of reproducing experiments in ML platforms and found that the platforms which have the most users have a relatively lower score in reproducibility. In this paper, we studied the reproducibility of pre-trained models across different bindings for the same ML framework.
 
To improve the reproducibility of ML models, many researchers have conducted studies to understand and resolve non-deterministic factors in ML software. Pham~\etal\cite{Pham_Problems_2020} studied nondeterminism-introducing-factors in ML frameworks~(e.g., weight initialization and parallel processes) and found that these factors can cause a 10\% accuracy difference in ML models. To improve the reproducibility of ML models, Chen~\etal\cite{chen_towards_2022} suggested using patching to minimize nondeterminism in hardware and proposed a record-and-reply approach to eliminate randomness in software. In addition, they provided guidelines for producing a reproducible ML model. Nagarajan~\etal\cite{Nagarajan_Reproducibility_2019} studied deterministic implementation for deep reinforcement learning and proposed a deterministic implementation of deep Q-learning by identifying and controlling five common sources of nondeterminism.

\subsection{Empirical Studies of ML Frameworks}
Many empirical studies of ML frameworks exist that study software quality aspects such as software bugs~\cite{Jia_bugs_2020, jia_bugs_2021, Chen_bugs_2022}, technical debt~\cite{sculley_hidden_2015, liu_debts_2020}, and programming issues~\cite{zhang_empirical_2019, humbatova_taxonomy_2020, islam_what_2019}. However, no prior work has investigated the impact of bindings for ML frameworks on the ML software quality.

Many studies have focused on the bugs of ML frameworks. Jia~\etal\cite{Jia_bugs_2020, jia_bugs_2021} investigated TensorFlow's GitHub repository and identified six symptoms and eleven root causes of bugs in TensorFlow. In addition, they found that most bugs are related to interfaces and algorithms. Chen~\etal\cite{Chen_bugs_2022} studied bugs from four ML frameworks and investigated the testing techniques in these frameworks. They showed that the most common root cause of the bugs is the incorrect implementation of algorithms, and the current testing techniques have a low percentage of test coverage.

ML software has ML-specific technical debts such as unstable data dependence, hidden feedback loop, and model configuration debts~\cite{sculley_hidden_2015}. This technical debt can hurt the maintainability of ML systems and introduce extra costs. Liu~\etal\cite{liu_debts_2020} analyzed self-admitted technical debt in 7 DL frameworks and concluded that technical debt is common in DL frameworks, although application developers are often unaware of its presence.

Researchers have also aimed to understand the ML frameworks from a developer perspective to study the programming issues when using an ML framework. They typically researched the questions and answers~(Q\&As) of developers about ML frameworks on Stack Overflow~(SO). Zhang~\etal\cite{zhang_empirical_2019} investigated Q\&As which are related to TensorFlow, PyTorch and Deeplearning4j on SO and reported that model migration is one of the most frequently asked questions. Humbatova~\etal\cite{humbatova_taxonomy_2020} studied Q\&As of these three ML frameworks on SO as well and included interviews with developers and researchers to build a taxonomy of faults in ML systems. Islam~\etal\cite{islam_what_2019} mined Q\&As about ten ML frameworks on SO and reported that developers need both static and dynamic analysis tools to help fix errors.


\subsection{FFIs and Bindings in Software Engineering}

FFIs and language bindings are instrumental in software engineering, serving as bridges that
enable different programming languages to collaborate seamlessly. These bridges often enable developers to develop applications in their language of choice while simultaneously using mature libraries that are developed in another language. The existing body of work predominantly proposes approaches to design and improve such bindings and FFIs within one specific language. For instance, Yallop~\etal\cite{jeremy_ffi_2018} conducted experiments to create bindings for using the \textit{ctypes} library in OCaml. Their study differentiated the performance of dynamic and static bindings, revealing that static bindings could be between 10 to 65 times faster than their dynamic counterparts. This finding aligns with our investigation into the time costs associated with diverse ML software bindings.

Researchers also proposed several approaches to FFIs. For instance, Bruni~\etal\cite{bruni_ffi_2013} introduced an FFI approach called NativeBoost. This approach requires minimal virtual machine modifications and generates native code directly at the language level. They compared the time cost of different FFIs and the results show that NativeBoost is competitive. Ekblad~\etal\cite{ekblad_ffi_2015} presented an FFI tailored for web-targeting Haskell dialects, emphasizing simplicity and automated marshalling. The authors compare their FFI with the vanilla FFI, which is based on C calling conventions, and show that their FFI has some advantages in terms of simplicity and expressiveness, safety, without introducing excessive performance~(i.e., time cost) overhead.

In addition, Ravitch~\etal\cite{ravitch_ffi_2009} automated the generation of library bindings using static analysis, aiming to simplify the often laborious manual creation process. Their method not only refined the automated binding generation but also unveiled type bugs in manually created bindings, highlighting potential threats to software correctness. Meanwhile, Grimmer~\cite{grimmer_ffi_2014} explored high-performance language interoperability in multi-language runtimes. Their approach leveraged just-in-time (JIT) compilers to optimize across language borders, enhancing the efficiency of cross-language operations.

To the best of our knowledge, our study is the first to systematically investigate the impact of using different language bindings on ML software quality. While Ravitch~\etal\cite{ravitch_ffi_2009} touched upon type correctness in bindings, the unique challenges posed by the inherently non-deterministic nature of ML software remain under-explored. Our work stands out as we specifically evaluate the impact of bindings on the correctness of ML software for model training and inference across different languages. In addition, The computationally intensive nature of ML software introduces unique challenges when assessing time costs, especially when relying on GPUs. While time cost is a widely used metric in the domain of FFIs and bindings, existing works do not explore its significance within the context of ML frameworks. Our research actively fills this void, presenting a comprehensive analysis of time costs associated with different bindings in ML software on CPUs and GPUs.

\section{Threats to Validity}\label{sec:threadstovalidity}



\subsection{Construct validity}
We use the accuracy metric to assess the correctness of TensorFlow and PyTorch bindings on model training and inference since it is a widely used metric among researchers and developers~\cite{elshawi_dlbench_2021, guo_empirical_2019, chirodea_comparison_2021, gevorkyan_review_2019, luo_comparison_2020}. However, other metrics may also be used to assess correctness and use of other metrics could potentially change our results. For evaluating the time cost of bindings on model training, we ran training experiments on the GPU since training DL models on CPU is time-consuming and developers usually train DL models on GPU. The results might be different from those obtained by measuring the time cost on CPU. 

\subsection{Internal validity}
When implementing the studied models in TensorFlow and PyTorch bindings, we used the same/similar interfaces to ensure that the structures of these models are consistent across bindings. However, bindings might have different implementations for these interfaces~(or have hidden bugs) that result in different structures in the built models. We saved the built models in bindings~(via parameters or serialization) and loaded them back into the default Python bindings for TensorFlow and PyTorch to examine whether the structures were the same. The verification results confirm that the produced models in bindings have the same structures. 

TensorFlow's JavaScript binding does not support training and inference for GRU with \enquote{reset\_after=True}. Hence, we set \enquote{reset\_after=False} in the training experiment of TensorFlow's JavaScript binding for GRU and performed inference with a GRU model that was trained with \enquote{reset\_after=False} in the default Python binding. This setup differs from other bindings, although it has no effect on the model's structure. We compared the results from the JavaScript binding to the results in the Python binding using \enquote{reset\_after=False}, and our findings still hold. Future studies should investigate how one can automatically confirm that the configurations of the bindings are exactly the same.



\subsection{External validity}
We focused on TensorFlow and PyTorch bindings in our work and the results of our study might not apply directly to other ML frameworks. One reason could be that other ML frameworks could have a different implementation and do not provide GPU support. Furthermore, the findings of our investigation may not be able to generalize to other models and datasets. Future studies should leverage our methodology to analyze bindings for other ML frameworks using different models and datasets.

Our analysis focused on small to medium-sized models that are widely adopted in real-world applications. However, the implications for large-scale models, particularly frontier ML models with billions or trillions of parameters, require further investigation. Future research should build on our work to examine how the observed differences might persist or change at this extreme scale.

\section{Conclusion}\label{sec:conclusion}

In this paper, we investigate the impact on ML software quality (correctness and time cost) of using bindings for ML frameworks for DL model training and inference. 
We conducted model training and model inference experiments on three CNN-based models and two RNN-based models in TensorFlow and PyTorch bindings written in four different programming languages. 
The most important findings of our study are:

\begin{itemize}
	\item When training models, bindings for ML frameworks can have various training accuracy curves and slightly different test accuracy values for the trained models.
	\item Bindings have different training times for the same model, and the default Python bindings for ML frameworks may not have the fastest training time.
	\item Bindings for ML frameworks have the capabilities to reproduce the accuracy of pre-trained models for inference.
	\item Bindings for ML frameworks have different inference times for the same pre-trained model and certain models in bindings on the CPU can outperform other bindings on the GPU.
\end{itemize}

Our findings show that developers can utilize a binding to speed up the training time for an ML model. For pre-trained models, developers can perform inference in their favoured programming language without sacrificing accuracy, or they can choose a binding that has better inference time. 



\section*{Disclaimer}
Any opinions, findings, and conclusions, or recommendations expressed in this material are those of the author(s) and do not reflect the views of Huawei.

\begin{acks}
The work described in this paper has been supported by the ECE-Huawei Research Initiative (HERI) at the University of Alberta.
\end{acks}

\bibliographystyle{ACM-Reference-Format}
\bibliography{main.bib}

%
%
%
%
%
%
%
%

\end{document}